\documentclass[a4paper]{article}
\usepackage{newunicodechar}
\usepackage{aas_macros}
\newunicodechar{∗}{\*}
\usepackage[T1]{fontenc}
\usepackage[utf8]{inputenc}
\usepackage{float}
\usepackage{varwidth}
\usepackage{amssymb}
\usepackage{geometry}
\usepackage[numbers,square, numbers, sort, compress]{natbib}

\makeatletter

\providecommand{\tabularnewline}{\\}
\newenvironment{cellvarwidth}[1][t]
    {\begin{varwidth}[#1]{\linewidth}}
    {\@finalstrut\@arstrutbox\end{varwidth}}

\usepackage{graphicx}  
\usepackage{amsmath}   
\usepackage{amssymb}   
\usepackage{bm} 
\usepackage{dcolumn}
\usepackage{color}
\usepackage{colortbl}
\usepackage{colordvi}
\usepackage{mathrsfs}
\usepackage{amsfonts}
\usepackage{varioref}
\usepackage{float}
\usepackage{graphicx}
\usepackage{ragged2e}

\usepackage{bibunits}
\defaultbibliographystyle{utphys1} 
\defaultbibliography{references} 
\RequirePackage[colorlinks,citecolor=blue,urlcolor=magenta,linkcolor=blue]{hyperref}
\input epsf
\usepackage{tablefootnote}
\usepackage{multirow}
\usepackage[T1]{fontenc}
\usepackage[utf8]{inputenc}
\usepackage{amsmath, amsfonts, amssymb, mathrsfs}
\usepackage{graphicx}
\graphicspath{{./Report/}}
\usepackage{enumitem}
\usepackage{comment}

\usepackage[noabbrev]{cleveref}
\crefname{equation}{eq.}{eq.}
\crefname{subsection}{sec.}{sec.}
\crefname{figure}{fig.}{fig.}
\crefname{subfigure}{fig.}{fig.}
\crefname{table}{tab.}{tab.}
\usepackage[utf8]{inputenc}
\usepackage{amsmath, amsfonts, amssymb}
\usepackage{subcaption}
\usepackage{float}
\usepackage{geometry}

\makeatother

\title{Investigating the interplay of the braneworld gravity and the plasma environment on the black hole shadow}

\author{Siddharth Kumar Sahoo \footnote{521ph1007@nitrkl.ac.in} \footnote{siddharth.math.physics@gmail.com}~$^{1}$ and Indrani Banerjee \footnote{banerjeein@nitrkl.ac.in}~$^{1}$\\
{\small{$^{1}$Department of Physics and Astronomy, National Institute of Technology, Rourkela, Odisha-769008, India}}}

\date{ }
\makeindex

\begin{document}
\maketitle
\begin{abstract}
We investigate the shadow of a rotating braneworld black hole in dispersive plasma environments and assess the potential of the Event Horizon Telescope (EHT) observations to constrain braneworld gravity. The spacetime around a rotating braneworld black hole is modelled by a Kerr-Newman-like metric determined by its mass $M$, spin $a$, and tidal charge $q$, which encodes the gravitational effects of the bulk spacetime. We consider both inhomogeneous and homogeneous plasma environments characterised by plasma parameters $\alpha_i$ ($i=1,2\text{ and }3$) to study light propagation and the interplay of the background spacetime and the plasma environment in influencing the shadow size and shape. We find that as the plasma density increases, inhomogeneous plasma environments decrease the shadow size, however homogeneous plasma enlarges it. On studying the effect due to the background spacetime, we find  that $q<0$ (negative tidal charge) increases the shadow diameter, while $q>0$ decreases it. Using the EHT measurements of M87* and Sgr A*, we constrain the $(q,\alpha_i)$ parameter space. The EHT data constrains the tidal charge in the range $-1.15 \lesssim q \lesssim 0.45$ for M87* and $-0.65 \lesssim q \lesssim 0.8$ for Sgr A*  in the low density plasma limit, which is indeed the case for M87* and Sgr A*. 
However, for black holes surrounded by high density plasma, the shadow size is governed both by the background geometry as well as by the plasma environment.  In such cases, joint constraints from plasma density estimates and observed shadow angular diameters can provide valuable insights into the underlying spacetime geometry.

\end{abstract}

\section{Introduction\label{Introduction}}
Black holes are  one of the key predictions of general relativity (GR) \citep{Einstein:1916vd,Penrose:1964wq, Hawking:1970zqf, Hawking:1966vg,Wald:1984rg}. They are also one of the  best candidates to test GR in the strong field regime \citep{Yagi:2016jml,Will:2014kxa,Berti:2015itd,Bambi:2015kza,Psaltis:2008bb}.  
The images of M87* \citep{EventHorizonTelescope:2019dse, EventHorizonTelescope:2019ths, EventHorizonTelescope:2019pgp, EventHorizonTelescope:2019ggy} and Sgr A* \citep{EventHorizonTelescope:2022wkp, EventHorizonTelescope:2022wok, EventHorizonTelescope:2022xqj, EventHorizonTelescope:2022exc} captured by the EHT collaboration mark significant milestones in confirming the predictions of GR  in the regime of strong gravitational fields. These developments have also opened new windows for precision tests of GR  where potential deviations from GR may be detectable \citep{Will:2014kxa, Berti:2015itd, Psaltis:2008bb}. While GR has been extensively tested, several  issues still remain unresolved. These include its inadequacy in explaining the dark sector \citep{Bekenstein1984ApJ...286....7B,Bosma1981AJ.....86.1791B,Bosma:1981zz,SupernovaSearchTeam:1998fmf, SupernovaCosmologyProject:1998vns,Romanowsky:2003qv}, the inevitability of singularities \citep{Penrose:1964wq, Hawking:1970zqf, Hawking:1966vg}, and the lack of compatibility with quantum theory \citep{Page:2004xp}.  Modified theories of gravity are an attempt to address these shortcomings of GR.

In this study, we examine shadow of black holes in braneworld gravity surrounded by plasma environments.  Braneworld gravity arises naturally in higher-dimensional models such as the Randall–Sundrum (RS) scenario, where our observable Universe is treated as a 3-brane embedded in a higher-dimensional bulk spacetime \citep{Randall:1999vf, Randall:1999ee,Akama:1982jy}. In these models, gravitational degrees of freedom propagate into the bulk, while standard model fields remain confined to the brane. The effective four-dimensional gravitational field equations on the brane include additional corrections induced by the five-dimensional Weyl tensor \citep{Shiromizu:1999wj}. These non-local effects manifest themselves as a tidal charge term in the black hole geometry \citep{Dadhich:2000am}.  The rotating braneworld black hole solution is a ``Kerr--Newman-like'' braneworld black hole obtained in \citep{Aliev:2005bi,Aliev:2006qp}. The corresponding spacetime describes a rotating black hole on the brane with mass $M$ and spin parameter $a$, along with a tidal charge parameter $q$ that encodes the influence of the extra dimension. Unlike the electric charge in the Kerr--Newman solution, the tidal charge $q$ arises from the gravitational field in the bulk and can take either positive or negative values. A negative tidal charge enlarges the event horizon and can enhance strong-field gravitational effects, thereby providing a potential observational signature of higher-dimensional gravity. Thus, the geometry is uniquely characterized by the parameters
\(
M ,  a \text{ and }  q \).
In contrast to standard General Relativity, where deviations in strong-field observables typically require exotic matter or quantum corrections, in the present case, the tidal charge  modifies the photon dynamics around the black hole due to the influence of higher dimensions. This makes rotating braneworld black holes a compelling candidate for testing extra-dimensional gravity using the Event Horizon Telescope (EHT) observations of M87* and Sgr A*.
{ The shadow of braneworld BHs in vacuum was first investigated by Amarilla \& Eiora \citep{Amarilla:2011fx} which was followed by subsequent works \cite{Eiroa:2017uuq,Liu:2024brf,Badia:2021kpk,Blaschke:2022pgf,Bohra:2023vls,Zhang:2021npt,Neves:2020doc,Hou:2021okc}. The present work investigates the role of the tidal charge and the plasma environment on the shadow of braneworld BHs.}

 {
When photons originating from distant astrophysical sources or the accretion disk surrounding the BH reach its vicinity, they experience strong gravitational deflection. 
If the photons are incident at the critical impact parameters, they move in spherical photon orbits in the photon region surrounding the BH. These spherical photon orbits are unstable and a slight perturbation can either cause the photon to plunge into the event horizon or escape to infinity. The bright ring surrounding the dark patch (the black hole shadow) is obtained by tracing back the light rays from the observer's position at infinity to the photon region of the black hole where unstable spherical photon orbits can exist. By projecting the unstable spherical photon orbits on the observer's sky the critical curve is obtained that separates the photons that escape to a distant observer from those that fall into the black hole
\citep{Gralla:2019xty,Teo:2003ltt, perlick2000ray, Perlick:2004tq}.
This central dark patch surrounded by a bright ring is referred to as the \textit{black hole shadow}}. The shadow of the Schwarzschild BH was first analyzed by Synge \citep{Synge:1966okc}, and later this study was  extended to the Kerr spacetime by Bardeen \citep{1974IAUS...64..132B}. 
Subsequent works explored shadows in the presence of plasma, first for spherically symmetric spacetimes \citep{Perlick:2015vta} and later for rotating BHs \citep{Perlick:2017fio}. 

Investigating the effect of plasma on the black hole shadow is important since astrophysical BHs are invariably surrounded by plasma-rich accretion flows \citep{Abramowicz:2011xu, Perlick:2017fio}. Plasma modifies photon propagation in a frequency-dependent manner, with significant impact in the radio regime \citep{Er:2013efa,Rogers:2015dla,Muhleman1966PhRvL..17..455M,Muhleman1970PhRvL..24.1377M}.
The study of plasma effects on radio signal propagation traces back to measurements involving the solar corona \citep{Muhleman1966PhRvL..17..455M,Muhleman1970PhRvL..24.1377M}. Since then, light deflection in plasma-filled Schwarzschild and Kerr spacetimes has been widely analyzed \citep{2000rofp.book.....P, Bisnovatyi-Kogan:2010flt, Tsupko:2013cqa, 2013ApSS.346..513M, Crisnejo:2018uyn, Crisnejo:2018ppm, Crisnejo:2019xtp, Crisnejo:2019ril}, alongside investigations of strong lensing \citep{Tsupko:2013cqa, Liu:2016eju, 2000rofp.book.....P, Bisnovatyi-Kogan:2010flt, 2013ApSS.346..513M, Crisnejo:2018uyn, Crisnejo:2018ppm, Crisnejo:2019xtp, Crisnejo:2019ril}. Recent studies have increasingly emphasised gravitational lensing \citep{Kumar:2022zky, Crisnejo:2022qlv, Bisnovatyi-Kogan:2022yzj, Tsupko:2019axo, Sun:2022ujt, Er_2020, Kala:2024fvg, Kala:2022uog} and black hole (BH) shadows in plasma environments \citep{Yan:2019etp, Chowdhuri:2020ipb, Badia:2021kpk, Badia:2022phg, Bezdekova:2022gib, Briozzo:2022mgg}. 
{ The EHT observations also give compelling evidence regarding the presence of plasma near M87* and Sgr A* \citep{EventHorizonTelescope:2019pgp,EventHorizonTelescope:2021srq,EventHorizonTelescope:2022urf,EventHorizonTelescope:2024hpu,EventHorizonTelescope:2024rju}. The EHT collaboration at present operates mainly in the 230 GHz band while the 345 GHz band is under active development. Both the observing frequencies are far above the plasma frequency. In 2021, the Event Horizon Telescope Collaboration presented the first resolved polarimetric image of M87* at 230 GHz \citep{EventHorizonTelescope:2021srq} while the first resolved linear and circular polarization images of Sgr A* at 230 GHz was published in 2024 \cite{EventHorizonTelescope:2024hpu}. The polarized ring structure and inferred Faraday rotation measures for Sgr A* are consistent with radiatively inefficient accretion flow models \citep{EventHorizonTelescope:2024hpu,EventHorizonTelescope:2024rju} while for M87*, polarization measurements are in agreement with magnetically arrested disk models \citep{EventHorizonTelescope:2021srq,EventHorizonTelescope:2024dhe}. Higher-frequency extensions (e.g.\ to 345~GHz) offer a promising path to further reduce optical-depth effects and probe deeper emission layers, but the existing 230~GHz data already provide robust and compelling constraints on the plasma environment through emission and polarimetric diagnostics. Hence it is worthwhile to investigate the effect of plasma on the black hole shadow. The present work which investigates the interplay between the tidal charge and the plasma environment on the black hole shadow is an effort in that direction.}

In \cref{Braneworld black hole} we give an overview of the braneworld model considered and the rotating braneworld black hole. We discuss light propagation in plasma in \cref{Overview of light propagation in pressureless non-magnetized plasma}. In \cref{Shadow of the braneworld black hole}, we derive the expression of shadow of a black hole as seen by an observer at some finite distance. We discuss the plasma environments considered in our work and their effect on the shadow of the braneworld black hole in \cref{Study of plasma profiles}. In \cref{EHT Observations of M87 and Sgr A}, we discuss our methodology and the results of obtaining constraints on braneworld gravity from the EHT observations of M87* and Sgr A*. In \cref{conclusion}, we give our concluding remarks. We used geometrized units ($G=c=1$)  during all theoretical calculations and switched to CGS when comparing with astrophysical observations. Unless specified we have scaled all distances with the gravitational radius $r_g=GM/c^2$ and masses with $M$, the mass of the black hole.  The signature of the metric throughout our work is mostly positive.
 \section{\label{Braneworld black hole} Braneworld black hole}
The rotating braneworld black hole solution  first reported in \citep{Aliev:2005bi,Aliev:2006qp} is a generalisation of a static black hole solution on a 3-brane in 5-D gravity in the Randall-Sundrum scenario \citep{Akama:1982jy,Dadhich:2000am}.  In braneworld gravity, the characteristic size $l$ of these extra dimensions can vastly exceed the Planck length $l_{\!P}\sim10^{-33}\,\mathrm{cm}$ as in the original Randall–Sundrum construction \citep{Randall:1999ee}.
It is therefore natural to anticipate that black holes formed by collapsing brane matter
will be predominantly confined to the brane, with only a modest portion of their
horizons extending into the bulk.
When matter residing on a 3‑brane undergoes gravitational collapse without angular
momentum, the resulting object should resemble a Schwarzschild black hole at
astrophysical distances so that the well‑tested predictions of General Relativity
remain intact. Chamblin \emph{et al.}\ \citep{Chamblin:1999by} examined such collapses within the
Randall-Sundrum framework (see also \citep{Garriga:1999bq,Chamblin:2000md,Giddings:2000mu}),
presenting a ``black‑string’’ geometry whose intersection with the brane reproduces the
familiar Schwarzschild metric.

Consider a five‑dimensional bulk with metric $g_{ab}$ ($a,b=0,1,2,3,4$) that contains a
3‑brane equipped with the induced metric $h_{\mu\nu}$.
In bulk coordinates $x^{a}$ the Einstein's equations read \citep{Aliev:2005bi,Aliev:2006qp}
\begin{equation}
{}^{(5)}G_{ab}= -\Lambda_{5}\,g_{ab}
   +\kappa_{5}^{2}\!\left({}^{(5)}T_{ab}
   +\sqrt{\frac{h}{g}}\;\tau_{ab}\,\delta(Z)\right),
\label{5DEinstein}
\end{equation}
where $\kappa_{5}^{2}=8\pi G_{5}$, $\Lambda_{5}$ is the bulk cosmological constant,
${}^{(5)}T_{ab}$ denotes bulk matter, and $\tau_{ab}$ is the brane stress‑energy tensor. To extract the effective four‑dimensional dynamics we foliate the bulk with a family
of timelike hypersurfaces $\Sigma_{Z}$ defined by a scalar function $Z(x^{a})$, choosing
the brane to be located at $Z=0$.
Introducing the unit normal $n_{a}=N\,\partial_{a}Z$ (with lapse $N$) and a set of tangent
vectors $e^{a}_{\mu}=\partial x^{a}/\partial y^{\mu}$ ($\mu=0,\dots,3$), the induced metric on
the brane becomes
\begin{equation}
h_{\mu\nu}=g_{ab}\,e^{a}_{\mu}e^{b}_{\nu}.
\label{inducedMetric}
\end{equation}
The bulk line element can then be written in an ADM‑like form
\begin{equation}
ds^{2}=h_{\mu\nu}\,dy^{\mu}dy^{\nu}
      +2N_{\mu}\,dy^{\mu}dZ
      +\bigl(N^{2}+N_{\mu}N^{\mu}\bigr)dZ^{2},
\label{ADMmetric}
\end{equation}
where $N^{\mu}$ is the shift vector.

Using the Gauss–Codazzi relations together with the Israel junction conditions
(assuming a $Z_{2}$ symmetry across the brane) yield the effective gravitational field equations
on the brane:
\begin{equation}
G_{\mu\nu}= -\Lambda\,h_{\mu\nu}
       +\kappa_{4}^{2}\tau_{\mu\nu}
       +\kappa_{5}^{4}S_{\mu\nu}
       -W_{\mu\nu}
       -3\kappa_{5}^{2}U_{\mu\nu}.
\label{braneEinstein}
\end{equation}
where
\begin{gather}
    U_{\mu\nu}=-\frac{1}{3}\left(^{(5)}T_{\mu\nu} -\frac{1}{4}\ h_{\mu\nu}h^{\lambda\sigma~(5)}T_{\lambda\sigma}\right)
\end{gather}
encodes the effect of the traceless part of the bulk energy-momentum tensor on the brane. Further,
\begin{itemize}
    \item $\Lambda$ and $\kappa_{4}^{2}$ are the effective four‑dimensional
          cosmological constant and gravitational coupling;
    \item $S_{\mu\nu}$ is quadratic in the brane stress tensor $\tau_{\mu\nu}$;
    \item $W_{\mu\nu}=A_{\mu\nu}-\tfrac14 h_{\mu\nu}A$ represents the non-local gravitational effects of the bulk onto the brane and is related to the ``electric''
          part of the bulk Riemann tensor $A_{\mu\nu}={}^{(5)}R_{abcd}n^{a}n^{c}e^{b}_{\mu}e^{d}_{\nu}$ and encodes its traceless projection onto the brane;
\end{itemize}

Explicitly,
\begin{align}
\Lambda &=\tfrac12\!\left(\Lambda_{5}
          +\tfrac16\kappa_{5}^{4}\lambda^{2}
          -\kappa_{5}^{2}P\right), ~~~~~\kappa_{4}^{2}=\tfrac16\kappa_{5}^{4}\lambda,\nonumber \\[4pt]
S_{\mu\nu}&=-\tfrac14\!\Bigl[
      \tau^{\sigma}{}_{\mu}\tau_{\sigma\nu}
      -\tfrac13 \tau \tau_{\mu\nu}
      -\tfrac12 h_{\mu\nu}\!\bigl(\tau_{\lambda\sigma}\tau^{\lambda\sigma}
      -\tfrac13 \tau^{2}\bigr)\Bigr],
      \label{6}
\end{align}
with $\lambda$ the brane tension and $P={}^{(5)}T_{ab}n^{a}n^{b}$ the bulk pressure normal to
the brane.

In the event the energy-momentum tensor on the bulk and in the brane is absent, the terms involving $\tau_{\mu\nu}$, $S_{\mu\nu}$ and $U_{\mu\nu}$ in \cref{braneEinstein} goes to zero. Also, from \cref{6}, by suitable fine-tuning, $\Lambda$ can be made to be small enough to match with the present day cosmological constant value. However, $W_{\mu\nu}$ which depends on the bulk geometry acts like a source term in the 4-d effective gravitational field equations. These equations have been solved previously in \citep{Aliev:2005bi,Aliev:2006qp} to obtain the rotating black hole solution on the brane given by, 
\begin{gather}
    \label{the metric}
    ds^2=-\Bigg(1-\frac{2 r M(r)}{\rho^2}\Bigg)dt^2-\frac{4 a r M(r)}{\rho^2}\sin^2\theta ~dt d\phi+\frac{\rho^2}{\Delta}dr^2+\rho^2 \,d\theta^2+\mathcal{A}\sin^2\theta \,d\phi^2
\end{gather}
where, $$M(r)=1-\frac{q}{2r}$$ $$\Delta=r^2+a^2-2 r M(r)$$  $$\rho^2=\Delta+2\ r M(r)-a^2 \sin^2\theta$$ and $$\mathcal{A}=\Delta+ 2\ rM(r)+\frac{2 a^2 M(r)r\sin^2\theta}{\rho^2}$$
The above metric is identical
in form to the Kerr–Newman solution in GR, but the tidal charge $q$ appearing  in \cref{the metric} is associated with the electric part of the bulk Weyl tensor. Unlike the electric charge appearing in  the Kerr-Newmann solution, the tidal charge can be negative, leading to stronger gravitational attraction than
in standard GR. While a positive charge can be attributed either to electric charge arising in the Kerr-Newman scenario or the tidal charge, a negative charge in \cref{the metric} has no analogue in GR and bears distinctive imprints of the extra-dimensional scenario.
The event horizon of the braneworld black hole is obtained by solving the equation $g^{rr}=0$ which gives us two solutions,
\begin{gather}
    \label{event horizon}
    r_{h{\pm}}=1\pm\sqrt{1-a^2-q}
\end{gather}
Note that, since $q$ can be negative the extremal black hole can have spin $a>1$. In what follows we will investigate the motion of photons in the above mentioned spacetime in presence of plasma environment.

\section{\label{Overview of light propagation in pressureless non-magnetized plasma}Overview of light propagation in pressureless, non-magnetized plasma}
We consider the case of non-magnetized, pressureless plasma in a stationary, axisymmetric spacetime.  
The plasma environment is characterized by the plasma frequency $\omega_p$ and the electron number density $n_e$. Both $\omega^2_p$ and $n_e$  are connected by the relation
\begin{gather}
    \label{plasma frequency electron number density relation}
    \omega^2_p=\frac{4\pi e^2}{m_e} n_e
\end{gather}
where $e$ and $m_e$ are the charge and mass of the electron, respectively.  In the absence of plasma ($\omega^2_p=0$), the path of the light rays follow the null geodesics  of the spacetime and can be obtained by solving the geodesic equations.  In the presence of plasma, light ray geodesics are no longer null geodesics because of the  dispersive effects of the plasma environment. 

 {In a non-magnetized, presureless  plasma environment,} the light ray geodesics can be computed by using the Hamiltonian formalism. The Hamiltonian $\mathcal{H}$ of the light ray travelling through the plasma is given by  \citep{Perlick:2015vta,Perlick:2017fio,breuer1980propagation,breuer1981propagation}. 
\begin{gather}
    \label{ Hamiltonian of the light ray travelling through the plasma1}
     \mathcal{H}=\frac{1}{2}(g^{\mu\nu}p_{\mu}p_{\nu}+\omega^2_p)
\end{gather}
here $p_\mu$ is the covariant momentum of the light ray.  In our case, both $p_t$ and $p_\phi$ are conserved as $g_{\mu\nu}$ and $\omega^2_p$ are independent of $t$ and $\phi$.  {Furthermore}, the light ray must satisfy the propagation condition in the plasma to reach an observer \citep{Perlick:2015vta,Perlick:2017fio,KumarSahoo:2025igt}. In order to arrive at the propagation condition, consider a timelike observer with  4-velocity $u^\mu$ ($u^\mu u_\mu=-1$ ). The momentum of the light ray can be decomposed in terms of components along the direction of $u^\mu$ and orthogonal to $u^\mu$ as,
\begin{gather}
    \label{ decomposed momentum of the light ray }
p^\mu=\omega(x) u^\mu+k^\mu
\end{gather}
where $k^\mu$ is the wave vector of light orthogonal to $u^\mu$ ($k^\mu u_\mu=0$) and $\omega(x)$ is the frequency of the light ray (which varies with the location of the observer). Using the conditions $u^\mu u_\mu=-1$ and $k^\mu u_\mu=0$, we can express the $\omega(x)$ as
\begin{gather}
    \label{light frequency interems of k and u}
    \omega( x)=-p_\mu u^\mu
\end{gather}
which leads to
\begin{gather}
    \label{k mu}
    k_\mu=p_\mu+p_\nu u^\nu u_\mu
\end{gather}
Using \cref{k mu} in \cref{ Hamiltonian of the light ray travelling through the plasma1} and using the property $\mathcal{H}=0$ \citep{Perlick:2017fio,breuer1980propagation,breuer1981propagation} for photons, we get the relation
\begin{gather}
    \label{k omegap omega relation}
    k_\mu k^\mu=\omega(x)^2-\omega^2_p(x)
\end{gather}
Further, from  the orthogonality of $k^\mu$ with $u^\mu$ ($k_\mu k^\mu\geq0$), for light ray propagation in plasma it is important  to satisfy the condition \citep{Perlick:2015vta,Perlick:2017fio},
\begin{gather}
    \label{omega omega p inequatlity}
    \omega^2(x)\geq \omega^2_p(x)
\end{gather}
The frequency of light $\omega(x)$ at any location with respect to a static, timelike observer ($u^\mu=\delta^\mu_t/\sqrt{-g_{tt}}$) can be calculated using the relation \citep{Perlick:2015vta,Perlick:2017fio},
\begin{gather}
    \label{omega x omeganot relation}
    \omega(x)=\frac{\omega_0}{\sqrt{-g_{tt}}}
\end{gather}
where $\omega_0=-p_t$. From \cref{omega x omeganot relation}, $\omega_0$ is the asymptotic frequency of the photon. Furthermore, the condition \cref{omega omega p inequatlity} can be used to obtain restrictions on the refractive index of the plasma $\mathbf{n}(r,\theta)$ given by
\begin{gather}
    \label{refractive index}
    \mathbf{n}(x)=\sqrt{1-\frac{\omega^2_p(x)}{\omega^2(x)}}
\end{gather}
For the light to propagate to a location $x$, $\mathbf{n}(x)\geq0$ \citep{Perlick:2017fio}.

\subsection{\label{The geodesics in presence of plasma satisfying separability condition}The geodesics of light rays in presence of plasma}

The Hamiltonian of the system in the presence of non-magnetized, pressureless plasma is given by,
\begin{gather}
    \label{hamiltonian general form}
    \mathcal{H}=\frac{1}{2}(g^{\mu\nu}p_{\mu}p_{\nu}+\omega^2_p)
\end{gather}
For a stationary, axisymmetric spacetime $\mathcal{H}$ is given by,
\begin{gather}
\label{expanded hamiltonian}
    \mathcal{H}=\frac{1}{2}(g^{tt}p^2_{t}+2g^{t\phi}p_{t}p_{\phi}+g^{\phi\phi}p^2_{\phi}+g^{rr}p^2_{r}+g^{\theta\theta}p^2_{\theta}+\omega^2_p)
\end{gather}
Substituting the metric components given in \cref{the metric} we get,
\begin{gather}
    \label{substitution of metric components in hamiltonian}
    -\Bigg(\frac{(\Delta+2\ rM(r))^2-a^2\Delta\sin^2\theta}{\rho^2\Delta}\Bigg) p^2_t-\frac{4 ar M(r)}{\rho^2\Delta}p_t p_\phi+\frac{\rho^2-2 r M(r)}{\rho^2\Delta\sin^2\theta} p^2_\phi+\frac{\Delta}{\rho^2} p^2_r+\frac{1}{\rho^2}p^2_\theta+\omega^2_p=2 \mathcal{H}
\end{gather}
and then simplifying, we get
\begin{gather}
    \label{Expanded and rearranged hamiltonian}
    \frac{-1}{\Delta}\{(\Delta+2\ rM(r))p_t+a p_\phi\}^2+\Delta p^2_r+p^2_\theta+\Big( a\sin\theta p_t+\frac{p_\phi}{\sin\theta}\Big)^2+\rho^2\omega^2_p=\rho^2\mathcal{H}
\end{gather}
The Hamilton-Jacobi equation for the system is given by,
\begin{gather}
\label{HJ equation}
    \mathcal{H}+\frac{\partial\mathcal{S}}{\partial\lambda}=0\\
\label{p mu from hj formula}
\text{where, }p_\mu=\frac{\partial\mathcal{S}}{\partial x^\mu}
\end{gather}
Since for light rays $\mathcal{H}=0$, thus $ \frac{\partial\mathcal{S}}{\partial\lambda}=0$. Using \cref{Expanded and rearranged hamiltonian} in \cref{HJ equation} we get
\begin{gather}
\label{HJ equation semi separable}
    \frac{-1}{\Delta}\{(\Delta+2\ rM(r))p_t+a p_\phi\}^2+\Delta p^2_r+p^2_\theta+\Big( a\sin\theta p_t+\frac{p_\phi}{\sin\theta}\Big)^2+\rho^2\omega^2_p=0
\end{gather}
Further we use the ansatz $$\mathcal{S}=p_t t+S_r(r)+S_\theta(\theta)+p_\phi \phi$$
which when used in \cref{HJ equation semi separable} gives,
\begin{gather}
\label{HJ equation with ansatz}
    \frac{-1}{\Delta}\{(\Delta+2\ rM(r))p_t+a p_\phi\}^2+\Delta \Bigg(\frac{d S_r(r)}{d r}\Bigg)^2+\Bigg(\frac{d S_\theta(\theta)}{d \theta}\Bigg)^2+\Big( a\sin\theta p_t+\frac{p_\phi}{\sin\theta}\Big)^2+\rho^2\omega^2_p=0
\end{gather}
The above equation is separable \citep{Bezdekova:2022gib,Perlick:2017fio,KumarSahoo:2025igt} in $r$ and $\theta$ \textit{iff} 
\begin{gather}
\label{separable plasma}
    \omega^2_p=\frac{f(r)+g(\theta)}{\rho^2}
\end{gather}
Using \cref{separable plasma} in \cref{HJ equation with ansatz} and rearranging, we get,
\begin{multline}
    \label{HJ equation with plasma separated}
    \Bigg(\frac{d S_\theta(\theta)}{d \theta}\Bigg)^2+\Big( a\sin\theta p_t+\frac{p_\phi}{\sin\theta}\Big)^2+g(\theta)=\\\frac{1}{\Delta}\{(\Delta+2\ rM(r))p_t+a p_\phi\}^2-\Delta \Bigg(\frac{d S_r(r)}{d r}\Bigg)^2-f(r)
\end{multline}
Clearly, the Hamilton-Jacobi equation is now separable. Let the separability constant be $\mathcal{Q}$\citep{Carter:1968rr,Perlick:2017fio,Bezdekova:2022gib}, then, we obtain
\begin{gather}
\label{HJ radial and theta geodesics}
        \Delta \Bigg(\frac{d S_r(r)}{d r}\Bigg)^2=-(\mathcal{Q}+f(r))+\frac{1}{\Delta}\{(\Delta+2\ rM(r))p_t+a p_\phi\}^2\\
        \Bigg(\frac{d S_\theta(\theta)}{d \theta}\Bigg)^2=\mathcal{Q}-g(\theta)-\Big( a\sin\theta\, p_t+\frac{p_\phi}{\sin\theta}\Big)^2
\end{gather}
or 
\begin{gather}
    \label{radial and theta equation}
    \rho^4 \dot{r}^2=-\Delta(\mathcal{Q}+f(r))+\{(\Delta+2\ rM(r))p_t+a p_\phi\}^2\\
    \rho^4\dot{\theta}^2=\mathcal{Q}-\Big( a\sin\theta\, p_t+\frac{p_\phi}{\sin\theta}\Big)^2-g(\theta)
\end{gather}
$\mathcal{H}$ does not explicitly depend on $t\text{ and }\phi$, this gives us two more equations.
\begin{gather}
\label{t and phi equation}
    \rho^2\dot{t}=\frac{-((\Delta+2\ rM(r))^2-a^2\Delta\sin^2\theta)p_t-2 a r M(r) p_\phi}{\Delta}\\
    \rho^2\dot{\phi}=\frac{(\rho^2-2 r M(r))p_\phi-2 a r M(r)\sin^2\theta\, p_t}{\Delta\sin^2\theta}
\end{gather}
Finally, we scale $f(r),\ g(\theta),\ \mathcal{Q}$ and $p_\phi$  in \cref{HJ radial and theta geodesics,t and phi equation} with  $-p_t$. Thus, the final geodesic equations are given by
\begin{gather}
\label{scaled geodesic equations}
    \frac{\rho^4 \dot{r}^2}{\omega_0^2}=-\Delta(\mathcal{\chi}+f(r))+\{-(\Delta+2\ rM(r))+a \eta\}^2={V}(r)\\
    \label{scaled geodesic equations theta}
    \frac{\rho^4\dot{\theta}^2}{\omega^2_0}=\mathcal{\chi}-\Big( -a\sin\theta\, +\frac{\eta}{\sin\theta}\Big)^2-g(\theta)=J(\theta)\\
     \frac{\rho^2\dot{t}}{\omega_0}=\frac{((\Delta+2\ rM(r))^2-a^2\Delta\sin^2\theta)-2 a r M(r) \eta}{\Delta}\\
  \frac{  \rho^2\dot{\phi}}{\omega_0}=\frac{(\rho^2-2 r M(r))\eta+2 a r M(r)\sin^2\theta}{\Delta\sin^2\theta}
\end{gather}
where, $f(r)\equiv \frac{f(r)}{\omega^2_0}$, $g(\theta)\equiv \frac{g(\theta)}{\omega^2_0}$. Additionally, $\chi= \frac{Q}{\omega_0^2}$ and $\eta=\frac{p_\phi}{\omega_0}$ are the impact parameters.

 {In order to calculate the shadow of braneworld black hole in the presence of plasma, it is important to calculate the radius of spherical photon orbits ($\dot{r}=0,\ddot{r}=0$)\citep{Teo:2003ltt,Perlick:2017fio}. } The condition for  spherical photon orbits can be expressed in terms of the radial geodesic equation as,
\begin{gather}
\label{spherical photon orbit conditions}
    \dot{r}=0\implies{V}(r)=0\\
    \ddot{r}=0\implies{\frac{dV(r)}{dr}}=0
\end{gather}
By solving the above equations for $\chi$ and $\eta$ we get
\begin{gather}
\label{general chi and eta}
  \chi=  \frac{\Delta (2\ r M'(r)+2\ M(r)+\Delta')^2} {\Delta '^2(r)}\left(1\pm \sqrt{1-\frac{\Delta
   ' f'(r)}{(2\ rM'(r)+2 M(r)+\Delta')^2}}\right)^2-f(r)\\
   a \eta = \frac{1}{\Delta'}\left((\Delta+2\ rM(r) )\Delta'+(2\ rM'(r)+2\ M(r)+\Delta')\Delta\left(1\pm\sqrt{1-\frac{\Delta'f'(r)}{(2\ rM'(r)+2\ M(r)+\Delta')^2}}\right)\right)
\end{gather}
For the case of braneworld metric $\Delta=r^2+a^2-2(r-\frac{q}{2})$ and $\Delta'=2(r-1)$, thus we obtain from \cref{general chi and eta}
\begin{gather}
\label{eta in terms spo}
   a\eta=\left(a^2+r^2 -\frac{r\Delta
   }{r-1}\right)\pm \frac{\Delta  r
   \sqrt{1-\frac{(r-1) f'(r)}{2 r^2}}}{r-1}\\
   \label{chi in terms spo}
   \chi=\frac{ \Delta  r^2 \left(1\pm \sqrt{1-\frac{(r-1)
   f'(r)}{2 r^2}}\right)^2}{(r-1)^2}-f(r)
\end{gather}
 {Above equations can be used to compute the constants of motion $\chi $ and $\eta$ for a spherical photon orbit. In the case of rotating black holes we have  a photon region instead of a single spherical photon orbit. The allowed values of radii which lie in the photon region can be determined using \cref{scaled geodesic equations theta}.  We substitute  \cref{eta in terms spo,chi in terms spo} in \cref{scaled geodesic equations theta} and impose the condition $J(\theta)\geq0$ to get the inequality as,}
    \begin{gather}
    \label{rearranged range for photon orbits}
    \chi(r_p)a^2\sin^2\theta-(a\eta(r_p)-a^2\sin^2\theta)-g(\theta)a^2\sin^2\theta\geq0
\end{gather}
The values of $r_p$ which satisfy the equality in the above inequality at the equatorial plane mark the boundaries of the photon region. From  inequality \eqref{rearranged range for photon orbits} we observe that the allowed values of $r_p$ depend on $\theta$ unless $a=0$ or $\theta=0$. When $a=0$ or $\theta=0$, inequality \eqref{rearranged range for photon orbits} becomes an equality and we get,
\begin{gather}
    \label{photon radius for circular shadow}
    a\eta(r_p)=0
\end{gather}
The allowed spherical photon orbits which contribute to the formation of the black hole shadow,are unstable in nature \citep{Teo:2003ltt,Perlick:2017fio}. Thus, in addition to \eqref{rearranged range for photon orbits} and \cref{photon radius for circular shadow}, in order to determine $r_p$ which contribute to shadow, the following condition must be used \citep{Teo:2003ltt,Perlick:2017fio},
\begin{gather}
    \label{photon orbit unstability condittion}
    \frac{d^2V(r)}{dr^2}\Bigg|_{r=r_p}\geq0
\end{gather}

\section{\label{Shadow of the braneworld black hole}Shadow of the braneworld black hole in presence of plasma}
We now proceed to derive the  equation of the shadow for a braneworld black hole in the presence of plasma. We will use the approach as described in \citep{Grenzebach:2015oea,Perlick:2017fio,Bezdekova:2022gib}. 
For an observer at inclination angle $\theta_i$  and located at a distance $d$ from the black hole, we define the following tetrad   
\begin{gather}
    \label{Tsupko's tetrad}
    e^\mu_{(0)}=\frac{1}{\rho\sqrt{\Delta}}\Bigg((r^2+a^2),0,0,a\Bigg)\Bigg|_{(d,\theta_i)}\\
    e^\mu_{(1)}=\frac{1}{\rho}\Bigg(0,0,1,0\Bigg)\Bigg|_{(d,\theta_i)}\\
    e^\mu_{(2)}=\frac{1}{\rho\sin\theta}\Bigg(-a\sin^2\theta,0,0,-1\Bigg)\Bigg|_{(d,\theta_i)}\\
    e^\mu_{(3)}=\frac{-\sqrt{\Delta}}{\rho}\Bigg(0,1,0,0\Bigg)\Bigg|_{(d,\theta_i)}
    \end{gather}
The above tetrad is orthonormal and  $e^\mu_0$ is the 4-velocity of the observer.   Let \(\Gamma^\mu(\lambda)\) be a null/photon geodesic with tangent
\begin{gather}
\label{coord-tangent}
\dot\Gamma^\mu(\lambda)=\dot t\,\partial_t+\dot r\,\partial_r+\dot\theta\,\partial_\theta+\dot\phi\,\partial_\phi,
\end{gather}
where overdot denotes \(d/d\lambda\). At the observer position the same tangent may be expanded in the tetrad basis as
\begin{gather}
\label{tetrad-tangent}
\dot\Gamma^\mu(\lambda)=-\Upsilon\, e^\mu_{(0)}+\beta\bigl(\sin\gamma\cos\delta\,e^\mu_{(1)}+\sin\gamma\sin\delta\,e^\mu_{(2)}+\cos\gamma\,e^\mu_{(3)}\bigr),
\end{gather}
with positive coefficients \(\Upsilon,\beta\).
Using \(g_{\mu\nu}\dot\Gamma^\mu\dot\Gamma^\nu=-\omega_P^2\) evaluated at \((d,\theta_i)\) yields
\begin{gather}
\label{Ubeta-omegaP}
\Upsilon^2-\beta^2=\omega_P^2(d,\theta_i).
\end{gather}
Projecting \(\dot\Gamma^\mu\) onto \(e_{(0)}^\mu\) (i.e. computing \(g_{\mu\nu}\dot\Gamma^\mu e^\nu_{(0)}\)) and comparing with \cref{tetrad-tangent} gives
\begin{gather}
\label{Upsilon-expression}
\Upsilon = g_{\mu\nu}\dot\Gamma^\mu e^\nu_{(0)}
= -\frac{r^2+a^2}{\rho\sqrt{\Delta}} + \frac{a\,\eta}{\rho\sqrt{\Delta}}\Big|_{(d,\theta_i)} .
\end{gather}
Substituting \eqref{Upsilon-expression} into \eqref{Ubeta-omegaP} yields
\begin{gather}
\label{beta-expression}
\beta=\sqrt{\frac{\bigl(a\,\eta-(r^2+a^2)\bigr)^2-\Delta\bigl(f(r)+g(\theta)\bigr)}{\rho^2\Delta}}\Big|_{(d,\theta_i)}.
\end{gather}
Matching the \(\partial_r\) components of \eqref{coord-tangent} and \eqref{tetrad-tangent} gives
\begin{gather}
\label{rdot-relation}
-\beta\cos\gamma\sqrt{\frac{\Delta}{\rho^2}}=\dot r.
\end{gather}
Using the radial geodesic equation for \(\dot r\) and solving for \(\sin\gamma\) yields
\begin{gather}
\label{sin-gamma-general}
\sin\gamma=\sqrt{\frac{\bigl(\mathcal{\chi}(r_p)-g(\theta)\bigr)\Delta}{\bigl(a\,\eta(r_p)-(r^2+a^2)\bigr)^2-\Delta\bigl(f(r)+g(\theta)\bigr)}}\Big|_{(d,\theta_i)}.
\end{gather}
Comparing the \(\partial_\phi\) components and using $\phi$ geodesic equation produces
\begin{gather}
\label{phidot-relation}
\dot\phi=-\frac{a\,\Upsilon}{\sqrt{\rho^2\Delta}}-\frac{\beta\sin\gamma\sin\delta}{\rho\sin\theta}.
\end{gather}
Combining this with the appropriate \(\dot t,\dot\phi\) relations and \eqref{sin-gamma-general} gives
\begin{gather}
\label{sin-delta-general}
\sin\delta=\frac{-\eta(r_p)+a\sin^2\theta}{\sin\theta\sqrt{\mathcal{\chi}(r_p)-g(\theta)}}\Big|_{(d,\theta_i)}.
\end{gather}
Evaluated at the observer \((d,\theta_i)\), the local sky angles are therefore
\begin{gather}
\label{sin-gamma}
\sin\gamma=\sqrt{\frac{\bigl(\mathcal{\chi}(r_p)-g(\theta)\bigr)\Delta}{\bigl(a\,\eta(r_p)-(r^2+a^2)\bigr)^2-\Delta\bigl(f(r)+g(\theta)\bigr)}}\Big|_{(d,\theta_i)},\\[6pt]
\label{sin-delta}
\sin\delta=\frac{-\eta(r_p)+a\sin^2\theta}{\sin\theta\sqrt{\mathcal{\chi}(r_p)-g(\theta)}}\Big|_{(d,\theta_i)}.
\end{gather}
The shadow boundary consists of directions that, when followed backward, asymptote to spherical photon orbits at radius \(r_p\). For each such \(r_p\), we compute \(\mathcal{\chi}(r_p)\) and \(\eta(r_p)\) from the conserved quantities (using \cref{eta in terms spo,chi in terms spo}), then obtain \(\gamma(r_p)\) and \(\delta(r_p)\) from \cref{sin-gamma} and \cref{sin-delta}. The allowed range of \(r_p\) is determined by the condition that \(\sin\delta=\pm1\) at the extrema,
\begin{gather}
\label{photon-range}
\sin\delta(r_{p_{\min/\max}})=\pm1,
\end{gather}
equivalently
\begin{gather}
\label{photon-range-condition}
\frac{-\eta(r_p)+a\sin^2\theta_i}{\sin\theta_i\sqrt{\mathcal{\chi}(r_p)-g(\theta_i)}}=\pm 1,
\end{gather}
which enforces that contributing photon orbits have turning points at \(\theta=\theta_i\).
For each \(r_p\) within \([r_{p_{\min}},r_{p_{\max}}]\) there is a unique \(\gamma\) from \eqref{sin-gamma} and two corresponding \(\delta\) values separated by \(\pi\). Projecting the celestial sphere stereographically onto the plane tangent at \(\gamma=0\) yields the Cartesian shadow coordinates (observer distance \(d\)):
\begin{gather}
\label{cartesian-x}
x(r_p)=-2d\,\tan\!\biggl(\frac{\gamma(r_p)}{2}\biggr)\sin\delta(r_p),\\[6pt]
\label{cartesian-y}
y(r_p)=-2d\,\tan\!\biggl(\frac{\gamma(r_p)}{2}\biggr)\cos\delta(r_p).
\end{gather}The angular coordinates (in radians) are \(X=x/d\) and \(Y=y/d\), giving
\begin{gather}
\label{angular-X}
X(r_p)=-2\,\tan\!\biggl(\frac{\gamma(r_p)}{2}\biggr)\sin\delta(r_p),\\[6pt]
\label{angular-Y}
Y(r_p)=-2\,\tan\!\biggl(\frac{\gamma(r_p)}{2}\biggr)\cos\delta(r_p).
\end{gather}
These angular functions \(X(r_p),Y(r_p)\) directly determine the shadow outline as seen by the observer at \((d,\theta_i)\)\citep{Ghezelbash:2012qn,Perlick:2017fio,KumarSahoo:2025igt}.
If $ r_{p_1} $ denotes the radius of the spherical photon orbit at the maximum angular height of the shadow from the horizontal, i.e., $ Y_{max} = Y(r_{p_1}) $, then the vertical angular diameter $ \Delta\Theta $ of the shadow can be calculated using the following formula:

\begin{gather}
\label{theoretical angular diameter}
    \Delta\Theta = 2 Y_{max}
\end{gather}
To determine $ r_{p_1} $, we solve the equation:

\begin{gather}
\label{photon orbit for maximum angular width}
    \frac{dY(r_p)}{dr_p} \Bigg|_{(r_p = r_{p_1})} = 0
\end{gather}
where $ r_{p_{min}} < r_{p_1} < r_{p_{max}} $. For a non-rotating black hole, we find $ r_{p_1} = r_s $, where $ r_s $ is obtained by solving \cref{photon radius for circular shadow}. We can also compute the Schwarzschild deviation parameter $\delta_{sh}$ using the angular diameter $\Delta\Theta$ as
\begin{gather}
    \label{sch deviation parameter}
    \delta_{sh}=\frac{\Delta\Theta}{\Delta\Theta_{Sch}}-1
\end{gather}
In above equation $\Delta\Theta_{Sch}=\frac{3\sqrt{3}GM}{c^2d}$ is the angular diameter  of shadow cast by a Schwarzschild black hole of mass $M$  and located at distance $d$ from the observer. $\Delta\Theta$ is the angular diameter of the braneworld black hole computed for a given $q,\alpha$ and $a$ using \cref{theoretical angular diameter}.    

\section{\label{Study of plasma profiles}Study of plasma profiles}
We now proceed to study the variation of different aspects of the shadow of braneworld black hole  for different plasma profiles. The bound on the plasma frequency $\omega^2_p$  can be determined using the light propagation in plasma \cref{omega omega p inequatlity}. If we redefine the $\omega^2_p$  as,
\begin{gather}
    \label{plasma frequency with bound}
    \omega^2_p=\omega^2_b \frac{{f}(r)+{g}(\theta)}{\rho^2}
\end{gather}
where the maximum value of $\omega_b$ is determined by the bound on $\omega_p$ using \cref{omega omega p inequatlity}. Scaling \cref{plasma frequency with bound} with the asymptotic frequency of the light ray $\omega_0$, we obtain
\begin{gather}
    \label{scaled plasma frequency}
    \frac{\omega^2_p}{\omega^2_0}=\alpha \frac{{f}(r)+{g}(\theta)}{\rho^2}
\end{gather}
In the above equation  $\alpha=\frac{\omega^2_b}{\omega^2_0}$ is the plasma parameter. The maximum possible value of $\alpha$ can be  determined using \cref{omega omega p inequatlity} which yields,
\begin{gather}
\label{alpha max}
\alpha_{max}=\text{min}(F(r,\theta))
\end{gather}
where,
\begin{gather}
    \label{F r theta}
    F(r,\theta)=-g_{tt}^{-1} (r,\theta)\frac{\rho^2}{{f}(r)+{g}(\theta)}
\end{gather}
We consider three plasma profiles motivated from \citep{Perlick:2017fio} in our work. The three plasma profiles are as follows,
\begin{flalign}
\label{profile 1}
\quad\ \quad\ \quad
\omega^2_p &= \alpha_1\frac{\sqrt{r}}{\rho^2} 
&& \text{$f(r)=\sqrt{r}$, $g(\theta)=0$} &
\\[5pt]
\label{profile 2}
\quad\ \quad\ \quad
\omega^2_p &= \alpha_2\frac{1+2 \sin^2\theta}{\rho^2} 
&& \text{$f(r)=0$, $g(\theta)=1+2 \sin^2\theta$} &
\\[5pt]
\label{profile 3}
\quad\ \quad\ \quad
\omega^2_p &= \alpha_3 
&& \text{$f(r)= r^2$, $g(\theta)= a^2 \cos^2\theta$} &
\end{flalign}

It is clear that the maximum of plasma parameter $\alpha_{max}$ for a given plasma profile will vary with braneworld charge $q$ and spin $a$. We now proceed to study the dependence of $\alpha_{max}$ on $q$ and $a$ for the plasma profiles in \cref{profile 1,profile 2,profile 3}. 
\subsection{\label{Analysis of plasma profile 1 and its effect  on shadow of braneworld black hole}Effect of inhomogeneous plasma profile 1 on the shadow of braneworld black hole}
\begin{figure}[h!]
    \centering
    
    \begin{subfigure}{0.7\textwidth}
        \centering
        \includegraphics[width=\linewidth]{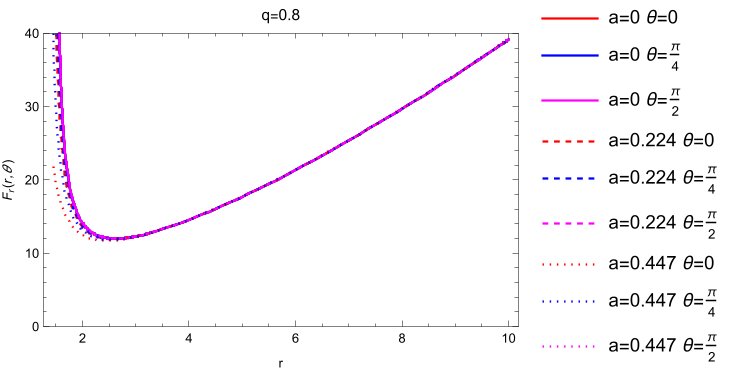}
        \caption{\label{Fr r theta study q 0.8}$q=0.8$}
    \end{subfigure}
    
     \vskip\baselineskip 
    
    \begin{subfigure}{0.7\textwidth}
        \centering
        \includegraphics[width=\linewidth]{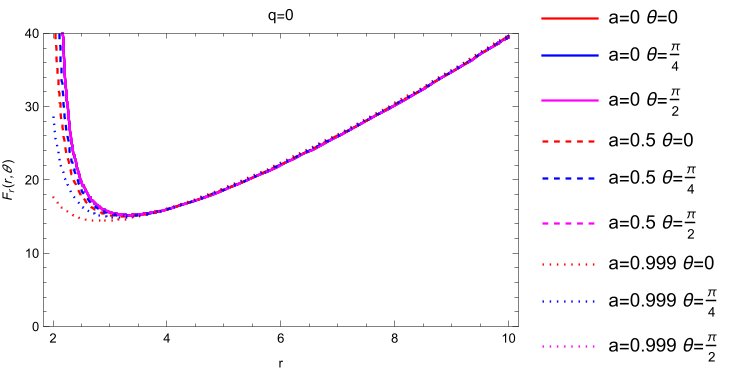}
        \caption{\label{Fr r theta study q 0}$q=0$}
    \end{subfigure}
    
     \vskip\baselineskip 
    
    \begin{subfigure}{0.7\textwidth}
        \centering
        \includegraphics[width=\linewidth]{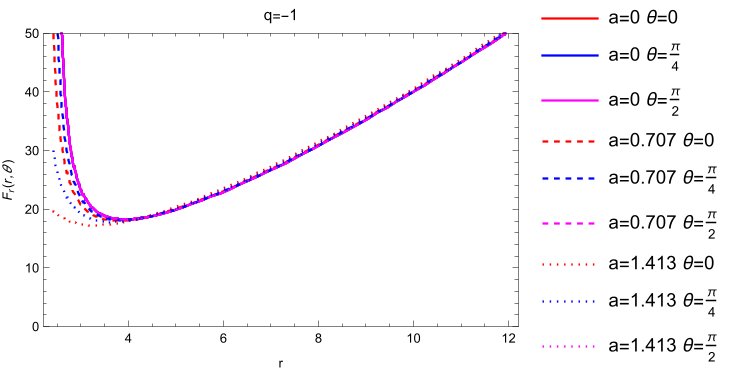}
        \caption{\label{F r theta study q -1}$q=-1$.}
    \end{subfigure}
  
    \caption{\label{Fr r theta study}Variation of $F_r(r,\theta)$ with distance for different values of $a$ and $\theta$ for profile 1.} 
\end{figure}

The plasma profile 1 in \cref{profile 1} is motivated from the Shapiro profile \citep{shapiro1974accretion,Perlick:2017fio} and at large distances the electron number density varies  as $n_e\sim r^{-3/2}$ representing a Bondi profile of spherical accretion. For plasma profile 1, we denote $F(r,\theta)$ by $F_r(r,\theta)$. We plot the variation of $F_r(r,\theta)$ with $r$ for $\theta=0, \pi/4$ and $\pi/2$. We do this for a fixed $q$ and $a=0, 0.5 a_{max}$ and $0.999a_{max}$, where $a_{max}=\sqrt{1-q}$ is the extremal spin for a given tidal charge $q$. In \cref{Fr r theta study}, the variation of $F_r(r,\theta)$ for the cases of tidal charge $q=0.8,0$ and $q=-1$ is plotted. We observe the following features of $F_r(r,\theta)$ 
\begin{itemize}
    \item The minima of $F_r(r, \theta)$ for a given $q$ always occurs for the near extremal spin and at angle $\theta=\pi/2$. This can be observed in \cref{Fr r theta study q 0.8,Fr r theta study q 0,F r theta study q -1} where the minima of $F_r(r,\theta)$ occurs for red dotted plot ($a=0.999 a_{max}$ and $\theta=\pi/2$).
    \item The minima of $F_r(r,\theta)$ decreases with increase in  $a$ for a given $q$. 
    \item From \cref{Fr r theta study q 0.8,Fr r theta study q 0,F r theta study q -1}, it can be clearly observed that the minima of $F_r(r,\theta)$ depends dominantly on the tidal charge $q$. 
In the case of positive tidal charge, the minima of $F_r(r,\theta)$ decreases as $q$ increases. On the contrary, the minima of $F_r(r,\theta)$ increases when $q$ becomes more and more negative.   
    \item $F_r(r,\theta)$ is insensitive to $a$ and $\theta$ as we go far from the horizon. However, near the horizon the effect of $a$ and $\theta$ are prominent. 
    
\end{itemize}

\begin{figure}[h!]
    \centering
    \begin{subfigure}{0.3\textwidth}
        \includegraphics[width=\linewidth]{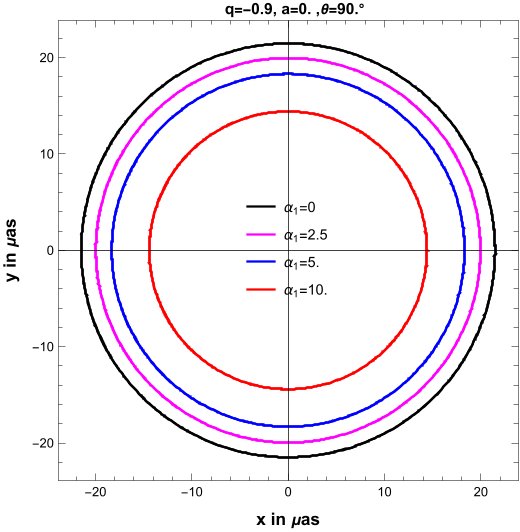}
        \caption{\label{image r_alpha_max_10._q_-0.9_a_0._theta_90}}
    \end{subfigure}
    \hfill
    \begin{subfigure}{0.3\textwidth}
        \includegraphics[width=\linewidth]{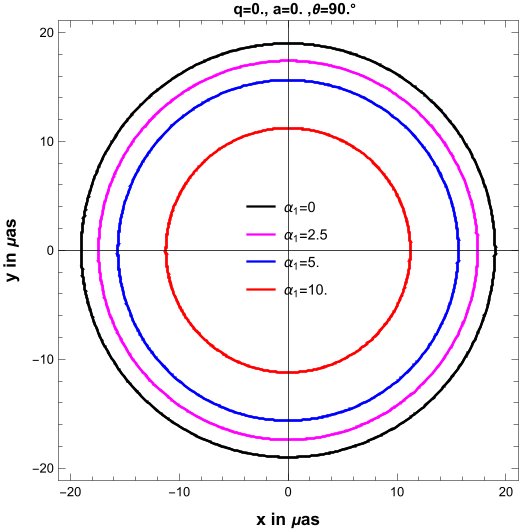}
        \caption{\label{image r_alpha_max_10._q_0._a_0._theta_90.}}
    \end{subfigure}
    \hfill
    \begin{subfigure}{0.3\textwidth}
        \includegraphics[width=\linewidth]{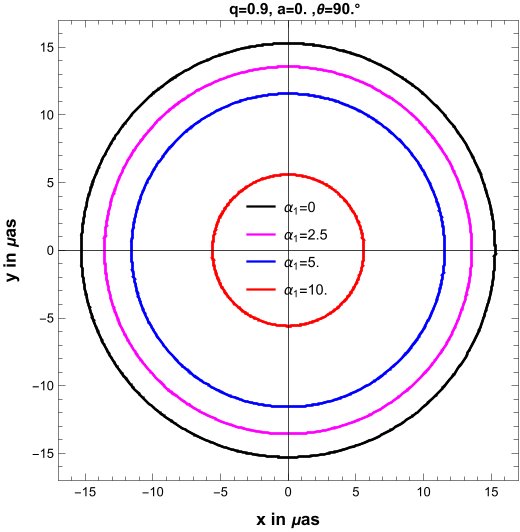}
        \caption{\label{image r_alpha_max_10._q_0.9_a_0._theta_90}}
    \end{subfigure}

    \vskip\baselineskip
    \begin{subfigure}{0.3\textwidth}
        \includegraphics[width=\linewidth]{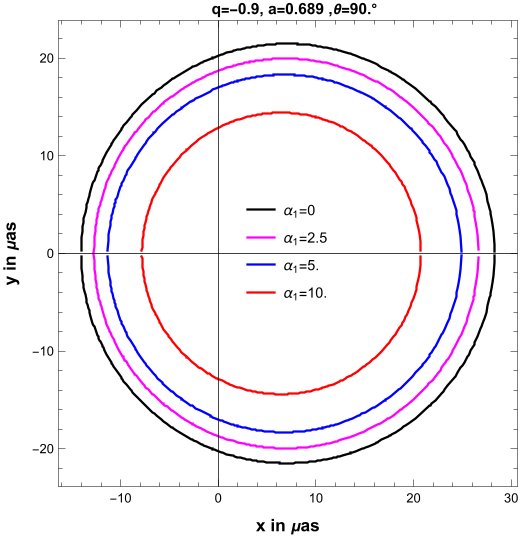}
        \caption{\label{image r_alpha_max_10._q_-0.9_a_0.689202_theta_90}}
    \end{subfigure}
    \hfill
    \begin{subfigure}{0.3\textwidth}
        \includegraphics[width=\linewidth]{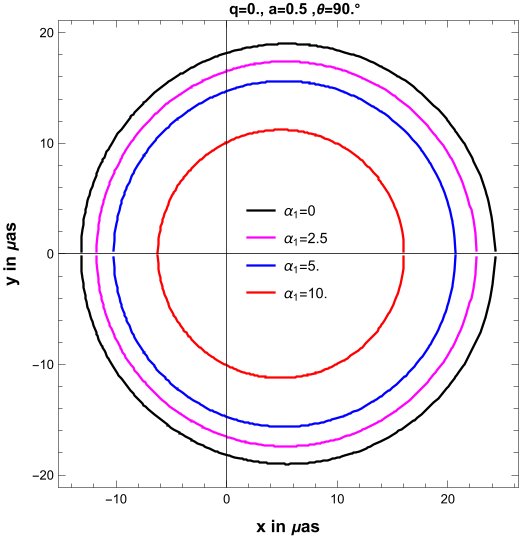}
        \caption{\label{image r_alpha_max_10._q_0._a_0.5_theta_90}}
    \end{subfigure}
    \hfill
    \begin{subfigure}{0.3\textwidth}
        \includegraphics[width=\linewidth]{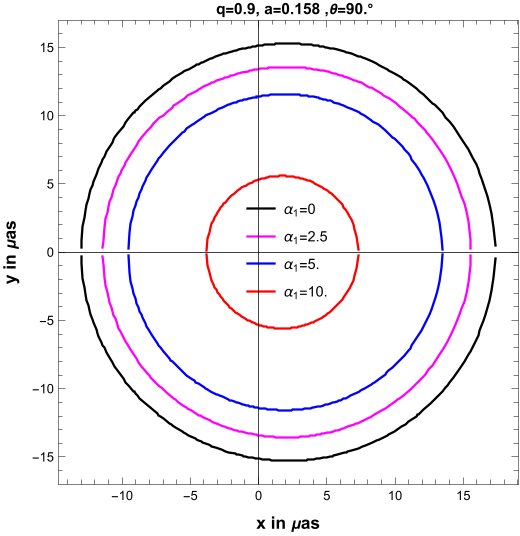}
        \caption{\label{image r_alpha_max_10._q_0.9_a_0.158114_theta_90}}
    \end{subfigure}

    \vskip\baselineskip
    \begin{subfigure}{0.3\textwidth}
        \includegraphics[width=\linewidth]{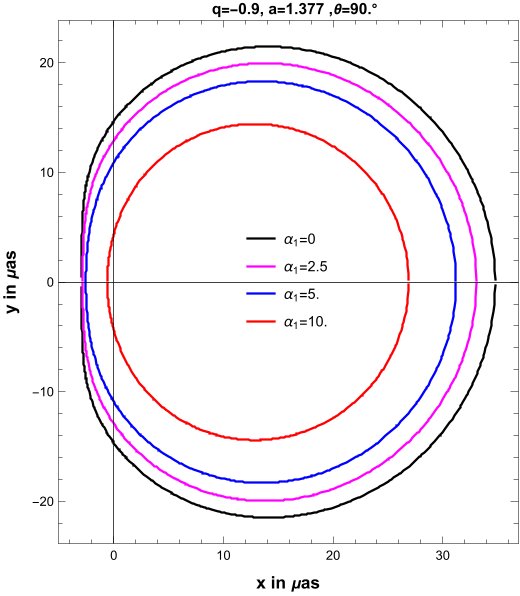}
        \caption{\label{image r_alpha_max_10._q_-0.9_a_1.37703_theta_90}}
    \end{subfigure}
    \hfill
    \begin{subfigure}{0.3\textwidth}
        \includegraphics[width=\linewidth]{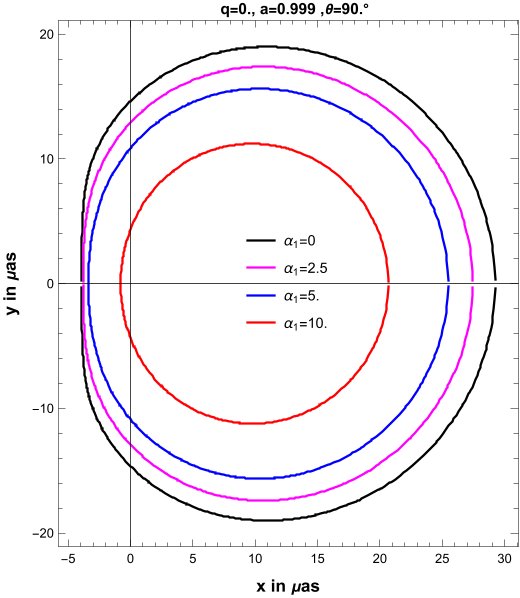}
        \caption{\label{image r_alpha_max_10._q_0._a_0.999_theta_90}}
    \end{subfigure}
    \hfill
    \begin{subfigure}{0.3\textwidth}
        \includegraphics[width=\linewidth]{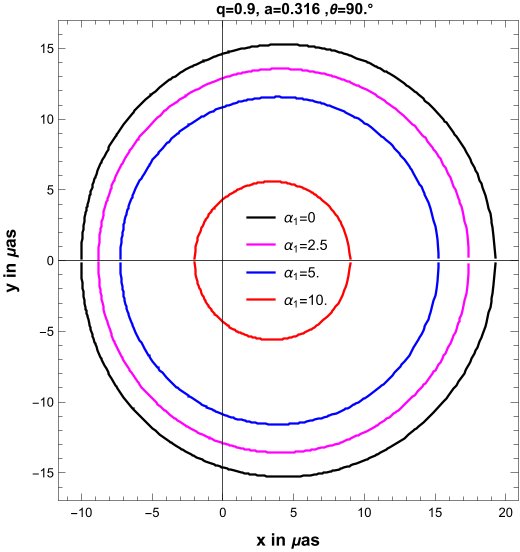}
        \caption{\label{image r_alpha_max_10._q_0.9_a_0.315912_theta_90}}
    \end{subfigure}

    \caption{\label{Plots showing effect of plasma profile 1 on shadow of braneworld black hole}Plots showing effect of plasma profile 1 on shadow of braneworld black hole for different combinations of $q$ and $a$ at $\theta=90^\circ$. }
\end{figure}

To analyze the impact of profile 1 on the shadow of a braneworld black hole, we plot the shadow for various combinations of $q$, $a$, and plasma parameter $\alpha_1$ in \cref{Plots showing effect of plasma profile 1 on shadow of braneworld black hole}. We set the mass $M = 6.2 \times 10^9 M_{\odot}$ and distance $d = 16.8$ Mpc of M87* as the parameters for the braneworld black hole. To understand the impact of the metric on the shape and size of the shadow we maintain the observation angle to be $\theta = 90^\circ$, since a combination of both high spin and high inclination gives rise to a deviation from circularity in the shape of the shadow.

In \cref{Plots showing effect of plasma profile 1 on shadow of braneworld black hole}, each plot illustrates the effect of varying the plasma parameter $\alpha_1$ for fixed $q$, $a$, and $\theta = 90^\circ$. The spin $a$ varies as $0$, $0.5 a_{\text{max}}$, and $0.999 a_{\text{max}}$ vertically down each column, while the tidal charge $q$ increases from $-0.9$ to $0$ and $0.9$ horizontally across each row. Thus, each plot reflects the effects of plasma parameter variation, with columns indicating spin variation and rows indicating tidal charge variation on the shadow. We make the following observations from \cref{Plots showing effect of plasma profile 1 on shadow of braneworld black hole}:

\begin{itemize}
    \item The size of the shadow  decreases with increase in plasma parameter $\alpha_1$, this can be clearly observed in all plots in \cref{Plots showing effect of plasma profile 1 on shadow of braneworld black hole}. 
    \item As the tidal charge $q$ increases the shadow size decreases and vice versa, for any given choice of $a$ and $\alpha_1$. In other words, shadow of a braneworld black hole will be smaller than that of a  Kerr black hole ($q=0$) when $q>0$ and will be larger when $q<0$.  This can be observed in each row in \cref{image r_alpha_max_10._q_-0.9_a_0._theta_90,image r_alpha_max_10._q_0._a_0._theta_90.,image r_alpha_max_10._q_0.9_a_0._theta_90}, \cref{image r_alpha_max_10._q_-0.9_a_0.689202_theta_90,image r_alpha_max_10._q_0._a_0.5_theta_90,image r_alpha_max_10._q_0.9_a_0.158114_theta_90} and \cref{image r_alpha_max_10._q_-0.9_a_1.37703_theta_90,image r_alpha_max_10._q_0._a_0.999_theta_90,image r_alpha_max_10._q_0.9_a_0.315912_theta_90}.
\item  At $\theta=90^\circ$ and in absence of plasma $\alpha_1=0$, the deviation from circular shape of the shadow increases with increase in spin $a$. This effect can be observed as we go down vertically in each column, i.e in  \cref{image r_alpha_max_10._q_-0.9_a_0._theta_90,image r_alpha_max_10._q_-0.9_a_0.689202_theta_90,image r_alpha_max_10._q_-0.9_a_1.37703_theta_90}, \cref{image r_alpha_max_10._q_0._a_0._theta_90.,image r_alpha_max_10._q_0._a_0.5_theta_90,image r_alpha_max_10._q_0._a_0.999_theta_90} and \cref{image r_alpha_max_10._q_0.9_a_0._theta_90,image r_alpha_max_10._q_0.9_a_0.158114_theta_90,image r_alpha_max_10._q_0.9_a_0.315912_theta_90}.   However, the deviation of circularity would be absent for $\theta=0^\circ$. Therefore, deviation from circularity is a combined effect of  $a$ and $\theta$.
\item   From  \cref{image r_alpha_max_10._q_-0.9_a_0._theta_90,image r_alpha_max_10._q_-0.9_a_0.689202_theta_90,image r_alpha_max_10._q_-0.9_a_1.37703_theta_90}; \cref{image r_alpha_max_10._q_0._a_0._theta_90.,image r_alpha_max_10._q_0._a_0.5_theta_90,image r_alpha_max_10._q_0._a_0.999_theta_90};  \cref{image r_alpha_max_10._q_0.9_a_0._theta_90,image r_alpha_max_10._q_0.9_a_0.158114_theta_90,image r_alpha_max_10._q_0.9_a_0.315912_theta_90}, we also observe that deviation from circularity decreases as $\alpha_1$ increases. Particularly, for sufficiently high $\alpha_1$ the shadow becomes nearly circular even for near extremal spin at $\theta=90^\circ$. Thus, the deviation from circularity of the shadow can also get affected by plasma environment, spin and angle of observation. This effect is also observed when the shadow is observed at different inclination angles, however the deviation from circularity is most prominent at $\theta=90^\circ$, hence we have not shown plots for other inclination angles.    

\end{itemize}


\subsection{\label{Analysis of plasma profile 2 and its effect  on shadow of braneworld black hole}Effect of inhomogeneous plasma profile 2 on the shadow of braneworld black hole}
\begin{figure}[h!]
    \centering
    
    \begin{subfigure}{0.7\textwidth}
        \centering
        \includegraphics[width=\linewidth]{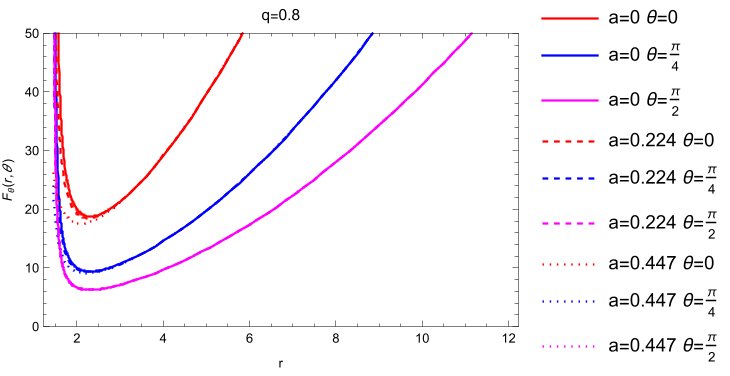}
        \caption{\label{Ftheta r theta study q 0.8}q=0.8}
    \end{subfigure}
    
     \vskip\baselineskip 
    
    \begin{subfigure}{0.7\textwidth}
        \centering
        \includegraphics[width=\linewidth]{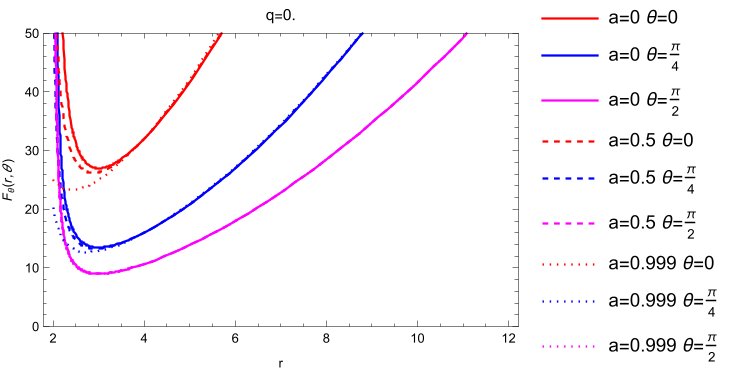}
        \caption{\label{Ftheta r theta study q 0}q=0}
    \end{subfigure}
    
     \vskip\baselineskip 
    
    \begin{subfigure}{0.7\textwidth}
        \centering
        \includegraphics[width=\linewidth]{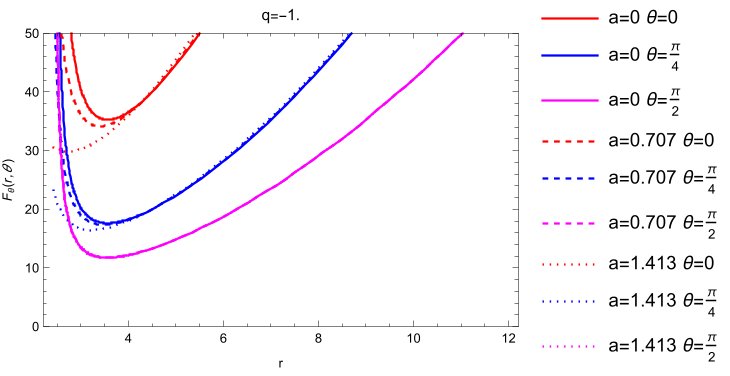}
        \caption{\label{Ftheta r theta study q -1}q=-1}
    \end{subfigure}
    
    \caption{\label{Ftheta r theta study}Variation of $F_\theta(r,\theta)$ with distance for different values of $a$ and $\theta$ for profile 2.} 
\end{figure}

The plasma profile 2 in \cref{profile 2} resembles a toroidal profile \citep{Rees:1982pe,Komissarov:2006nz,Mosallanezhad:2013gre}.  The $F(r,\theta)$ for profile 2 is represented by $F_\theta(r,\theta)$. The variation of $F_\theta(r,\theta)$ with $r$ for different $a$ and $\theta$, for a given $q$  are shown in \cref{Ftheta r theta study} . The minima of $F_\theta(r,\theta)$ correspond to the bound on plasma parameter $\alpha_2$. From \cref{Ftheta r theta study} we observe the following:

\begin{itemize}
    \item From \cref{Ftheta r theta study q 0.8,Ftheta r theta study q 0,Ftheta r theta study q -1} it is evident that $F_\theta(r,\theta)$ is sensitive to $\theta$ such that, the minima of $F_\theta(r,\theta)$ decreases with increase in $\theta$. However, the dependence on  $a$ is only prominent at distances close to the BH.

    \item  For a given inclination, the minima of  $F_\theta(r,\theta)$ decreases with increase in spin.
    \item The minima of $F_\theta(r,\theta)$ decreases with increase in positive $q$ and increases as $q$ becomes more and more negative as seen from \cref{Ftheta r theta study q 0.8,Ftheta r theta study q 0,Ftheta r theta study q -1}. This feature was also observed  in the $F_r(r,\theta)$ in case of profile 1.
    \end{itemize}

\begin{figure}[h!]
    \centering
    \begin{subfigure}{0.3\textwidth}
        \includegraphics[width=\linewidth]{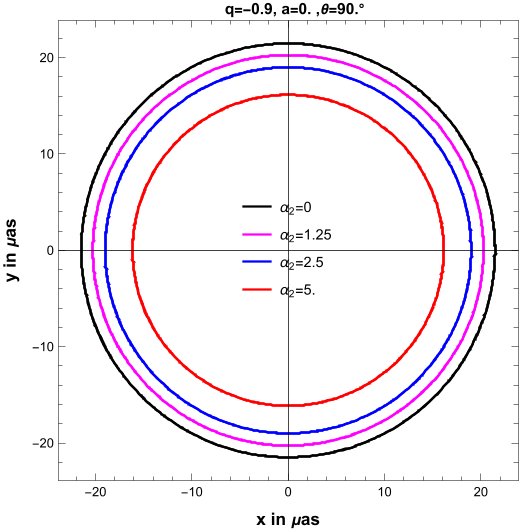}
        \caption{\label{theta_alpha_max_5._q_-0.9_a_0._theta_90}}
    \end{subfigure}
    \hfill
    \begin{subfigure}{0.3\textwidth}
        \includegraphics[width=\linewidth]{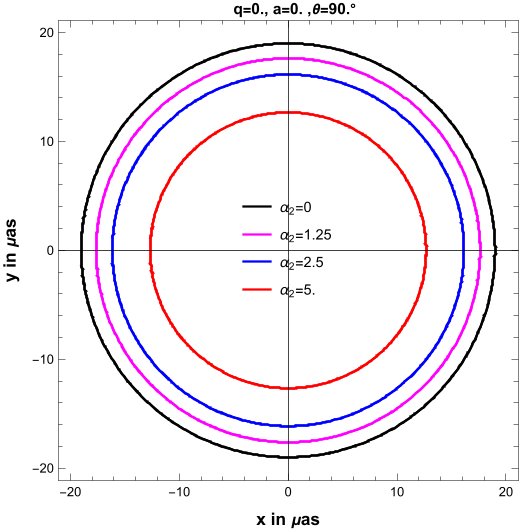}
        \caption{\label{theta_alpha_max_5._q_0._a_0._theta_90}}
    \end{subfigure}
    \hfill
    \begin{subfigure}{0.3\textwidth}
        \includegraphics[width=\linewidth]{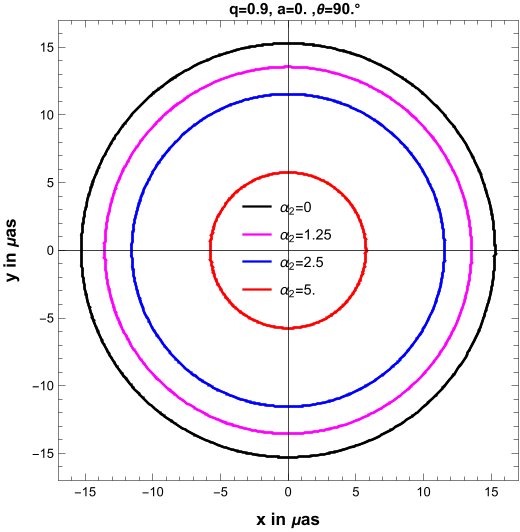}
        \caption{\label{image theta_alpha_max_5._q_0.9_a_0._theta_90}}
    \end{subfigure}

    \vskip\baselineskip
    \begin{subfigure}{0.3\textwidth}
        \includegraphics[width=\linewidth]{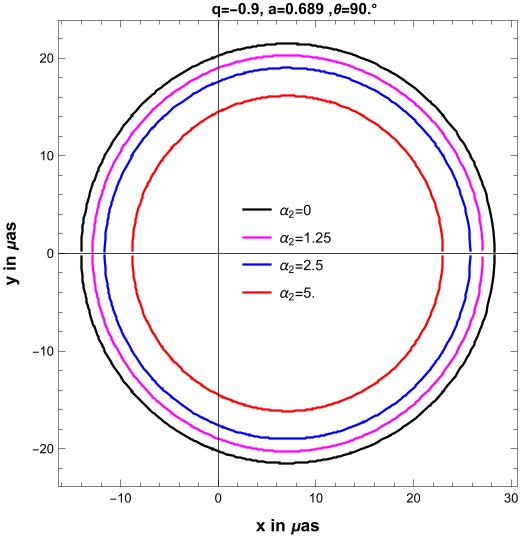}
        \caption{\label{image theta_alpha_max_5._q_-0.9_a_0.689202_theta_90}}
    \end{subfigure}
    \hfill
    \begin{subfigure}{0.3\textwidth}
        \includegraphics[width=\linewidth]{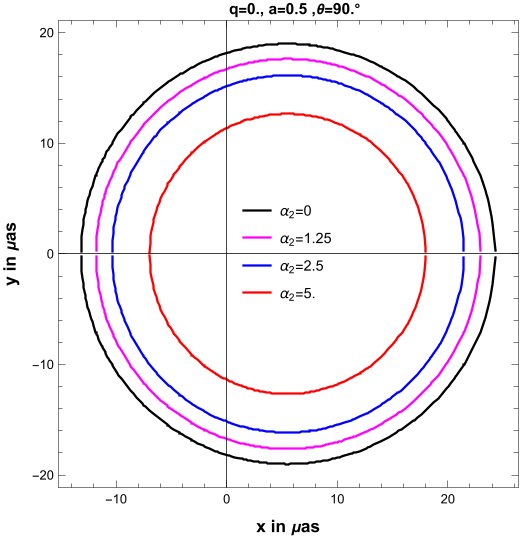}
        \caption{\label{image theta_alpha_max_5._q_0._a_0.5_theta_90}}
    \end{subfigure}
    \hfill
    \begin{subfigure}{0.3\textwidth}
        \includegraphics[width=\linewidth]{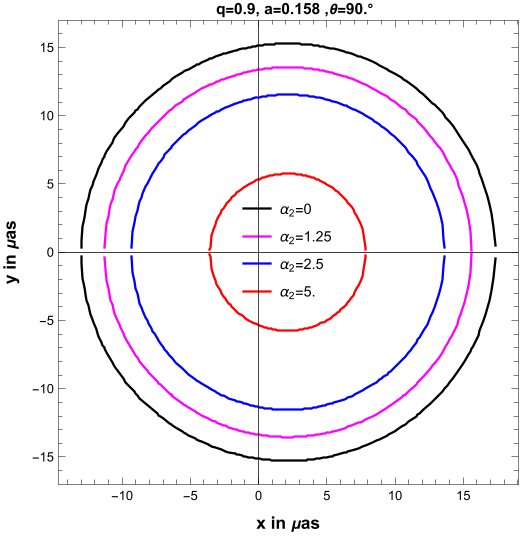}
        \caption{\label{image theta_alpha_max_5._q_0.9_a_0.158114_theta_90}}
    \end{subfigure}

    \vskip\baselineskip
    \begin{subfigure}{0.3\textwidth}
        \includegraphics[width=\linewidth]{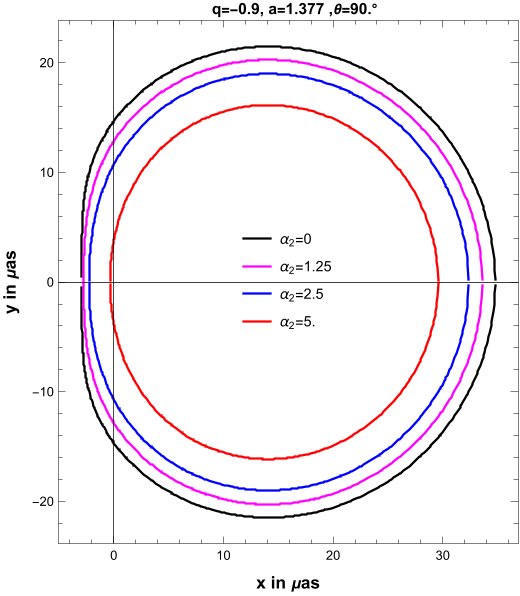}
        \caption{\label{image theta_alpha_max_5._q_-0.9_a_1.37703_theta_90}}
    \end{subfigure}
    \hfill
    \begin{subfigure}{0.3\textwidth}
        \includegraphics[width=\linewidth]{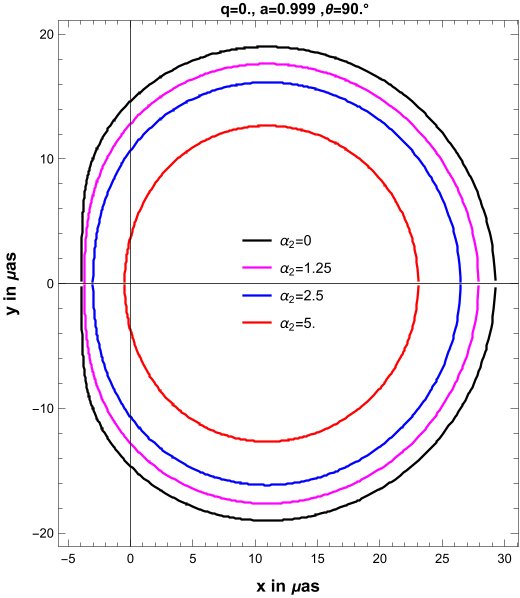}
        \caption{\label{image theta_alpha_max_5._q_0._a_0.999_theta_90}}
    \end{subfigure}
    \hfill
    \begin{subfigure}{0.3\textwidth}
        \includegraphics[width=\linewidth]{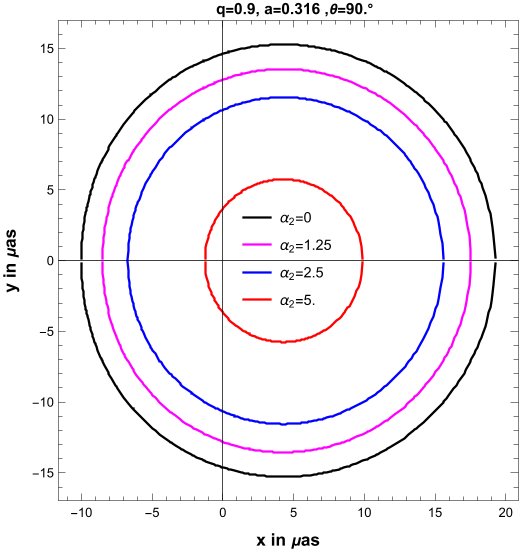}
        \caption{\label{image theta_alpha_max_5._q_0.9_a_0.315912_theta_90}}
    \end{subfigure}

    \caption{\label{Plots showing effect of plasma profile 2 on shadow of braneworld black hole}Plots showing effect of plasma profile 2 on shadow of braneworld black hole for different combinations of $q$ and $a$ at $\theta=90^\circ$. }
\end{figure}

Next, we analyze the effects of plasma profile 2 on braneworld black hole shadows. Using the approach detailed in \cref{Analysis of plasma profile 1 and its effect on shadow of braneworld black hole} for \cref{Plots showing effect of plasma profile 1 on shadow of braneworld black hole}, we plot the shadow for varying $q, a$ and $\alpha_2$, with $\theta$ fixed at $90^\circ$ in \cref{Plots showing effect of plasma profile 2 on shadow of braneworld black hole}. Our observations are as follows:

\begin{itemize}
    \item  The increase in shadow size with decreasing $q$ and the increase in  deviation from circularity with increasing $a$ (at $\theta=90^\circ$) as observed in \cref{Plots showing effect of plasma profile 1 on shadow of braneworld black hole} is also evident in \cref{Plots showing effect of plasma profile 2 on shadow of braneworld black hole}. 
    \item The shadow size decreases with increasing $\alpha_2$, and as we move downwards in each column of \cref{Plots showing effect of plasma profile 2 on shadow of braneworld black hole}, we observe a decrease in the deviation in circularity of the shadow with increasing $\alpha_2$, independent of the values of $a$ and $q$.
    \item Furthermore, comparing shadows for $\alpha_2\simeq \alpha_1\simeq 5$ in \cref{Plots showing effect of plasma profile 1 on shadow of braneworld black hole,Plots showing effect of plasma profile 2 on shadow of braneworld black hole} reveal that the contraction effect of plasma profile 2 is higher than that of plasma profile 1. This however is true at high inclination angles (e.g. $\theta_i\sim 90^\circ$) and the trend may change when the inclination angle is low, e.g. when $\theta_i\sim 17^\circ$ then profile 1 has marginally higher contracting effect compared to profile 2. Further, the contracting effect due to the inhomogeneous plasma profiles also depend on the distance-to-mass ratio. Hence, the trend may alter for different black hole sources. 
   
\end{itemize}

\begin{figure}[h!]
    \centering
    
    \begin{subfigure}{0.7\textwidth}
        \centering
        \includegraphics[width=\linewidth]{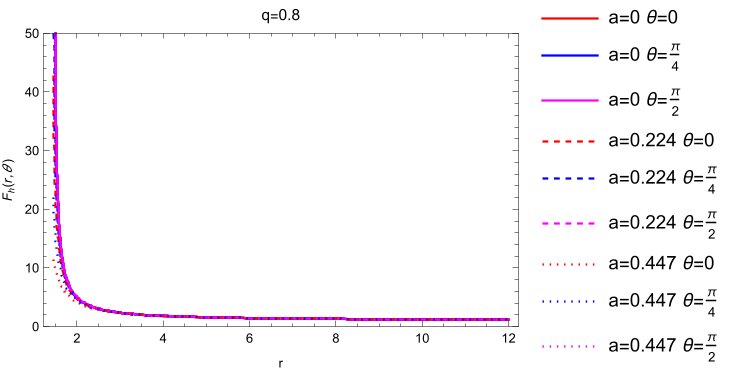}
        \caption{\label{Fh r theta study q 0.8}q=0.8}
    \end{subfigure}
    
     \vskip\baselineskip 
    
    \begin{subfigure}{0.7\textwidth}
        \centering
        \includegraphics[width=\linewidth]{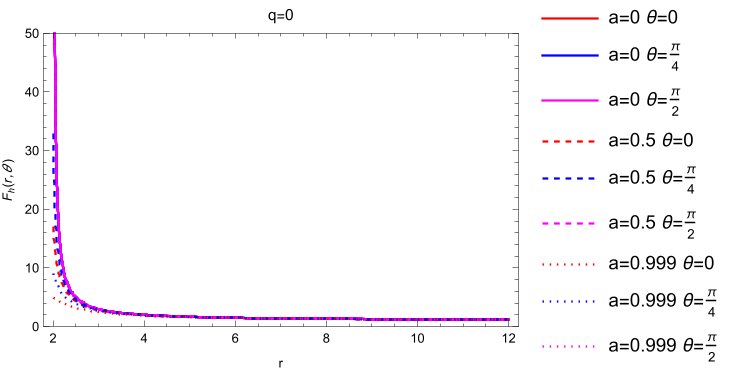}
        \caption{\label{Fh r theta study q 0}q=0}
    \end{subfigure}
    
     \vskip\baselineskip 
    
    \begin{subfigure}{0.7\textwidth}
        \centering
        \includegraphics[width=\linewidth]{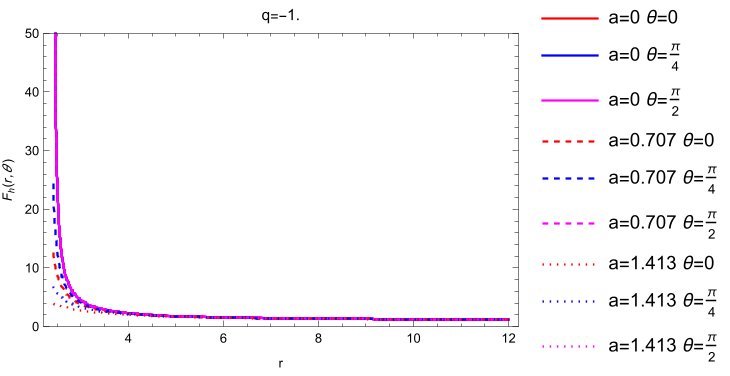}
        \caption{\label{Fh r theta study q -1}q=-1}
    \end{subfigure}
    
    \caption{\label{Fh r theta study}Variation of $F_h(r,\theta)$ with distance for different values of $a$ and $\theta$ for homogeneous plasma.} 
\end{figure}



\begin{figure}[t!]
    \centering
    \begin{subfigure}{0.3\textwidth}
        \includegraphics[width=\linewidth]{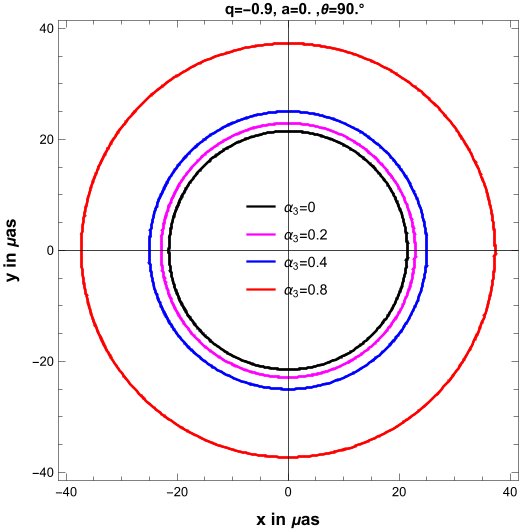}
        \caption{\label{image h_alpha_max_0.8_q_-0.9_a_0._theta_90}}
    \end{subfigure}
    \hfill
    \begin{subfigure}{0.3\textwidth}
        \includegraphics[width=\linewidth]{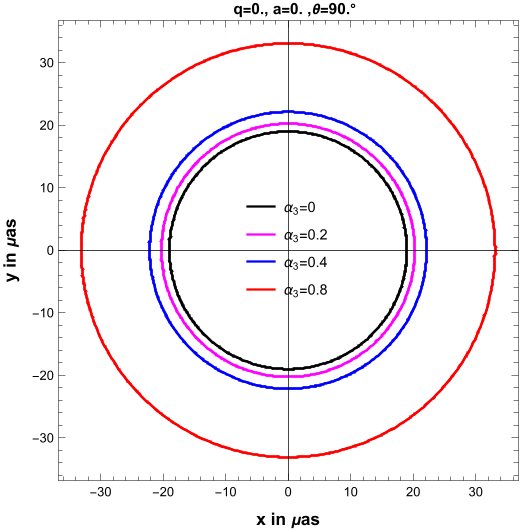}
        \caption{\label{image h_alpha_max_0.8_q_0._a_0._theta_90}}
    \end{subfigure}
    \hfill
    \begin{subfigure}{0.3\textwidth}
        \includegraphics[width=\linewidth]{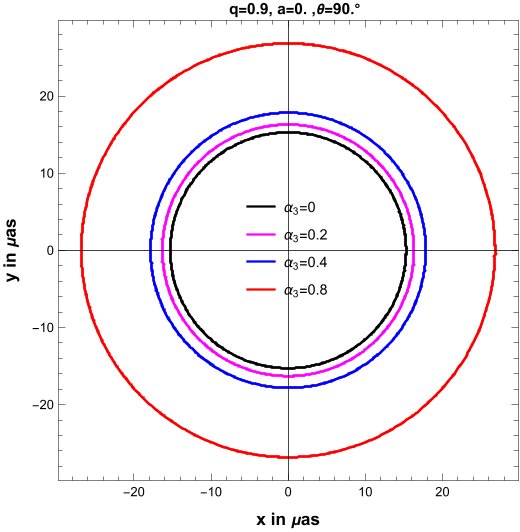}
        \caption{\label{image h_alpha_max_0.8_q_0.9_a_0._theta_90}}
    \end{subfigure}

    \vskip\baselineskip
    \begin{subfigure}{0.3\textwidth}
        \includegraphics[width=\linewidth]{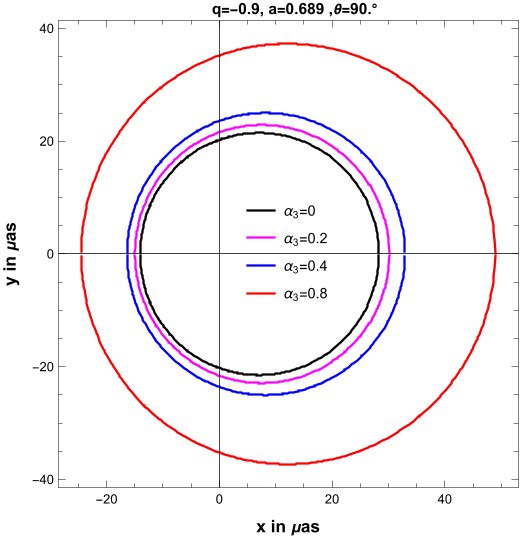}
        \caption{\label{image h_alpha_max_0.8_q_-0.9_a_0.689202_theta_90}}
    \end{subfigure}
    \hfill
    \begin{subfigure}{0.3\textwidth}
        \includegraphics[width=\linewidth]{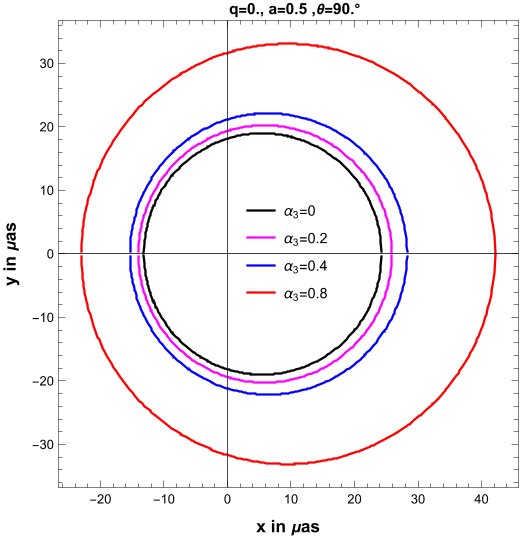}
        \caption{\label{image h_alpha_max_0.8_q_0._a_0.5_theta_90}}
    \end{subfigure}
    \hfill
    \begin{subfigure}{0.3\textwidth}
        \includegraphics[width=\linewidth]{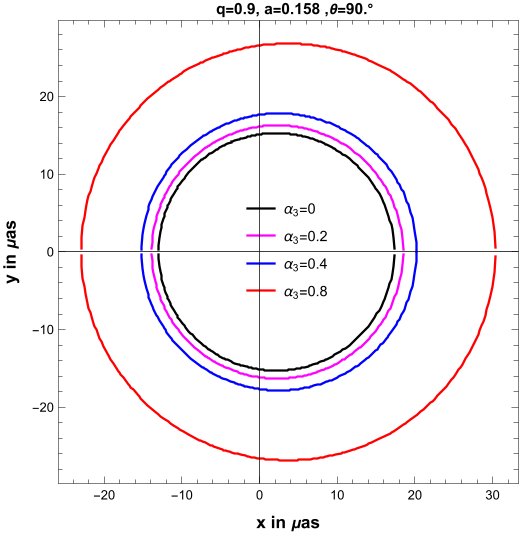}
        \caption{\label{image h_alpha_max_0.8_q_0.9_a_0.158114_theta_90}}
    \end{subfigure}

    \vskip\baselineskip
    \begin{subfigure}{0.3\textwidth}
        \includegraphics[width=\linewidth]{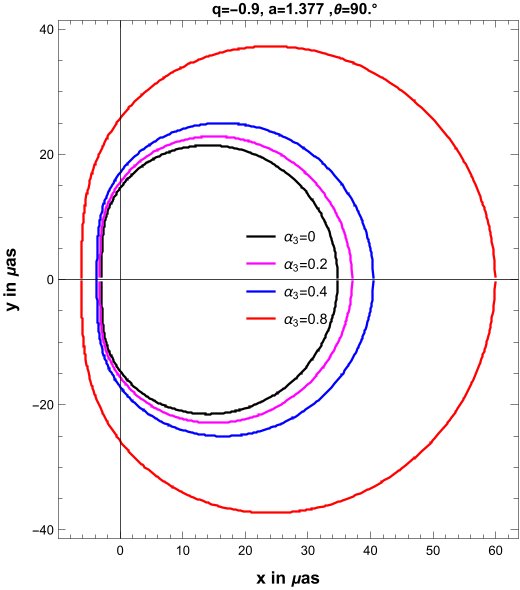}
        \caption{\label{image h_alpha_max_0.8_q_-0.9_a_1.37703_theta_90}}
    \end{subfigure}
    \hfill
    \begin{subfigure}{0.3\textwidth}
        \includegraphics[width=\linewidth]{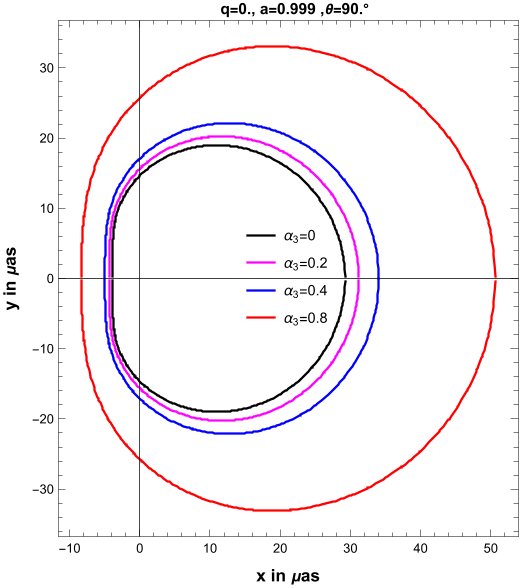}
        \caption{\label{image h_alpha_max_0.8_q_0._a_0.999_theta_90}}
    \end{subfigure}
    \hfill
    \begin{subfigure}{0.3\textwidth}
        \includegraphics[width=\linewidth]{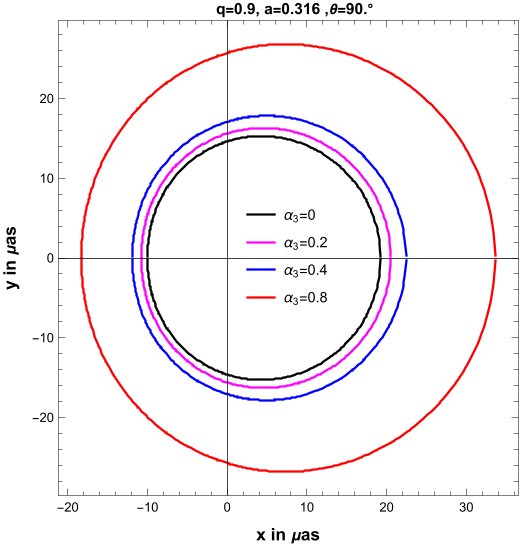}
        \caption{\label{h_alpha_max_0.8_q_0.9_a_0.315912_theta_90}}
    \end{subfigure}

     \caption{\label{Plots showing effect of homogeneous plasma  on shadow of braneworld black hole}Plots showing effect of homogeneous plasma on shadow of braneworld black hole for different combinations of $q$ and $a$ at $\theta=90^\circ$. }
\end{figure}

\subsection{Effect of homogeneous plasma on the shadow of braneworld black hole \label{Analysis of homogeneous plasma  and its effect  on shadow of braneworld black hole}}
For the sake of completeness, we also consider the homogeneous plasma profile given in \cref{profile 3}. In a homogeneous plasma, the electron density $n_e$ is constant. Using the condition $\mathbf{n}(r,\theta)\geq0$ (refer  \cref{refractive index}) we note that   $\alpha_{3max}\sim1$.   We can visualise this by plotting the variation of $F_h(r,\theta)$  ($F(r,\theta)$ for homogeneous profile) with $r$ for different $q$.  The plots for $F_h(r,\theta)$ variation for different $q$ are shown in \cref{Fh r theta study q 0.8,Fh r theta study q 0,Fh r theta study q -1}.   From \cref{Fh r theta study},  we draw the following conclusions:
\begin{itemize}
\item  $F_h(r,\theta)$ monotonically decreases with increase in distance from the BH and asymptotically approaches  to 1 for any $q$, $a$ and $\theta$ as seen in \cref{Fh r theta study q 0.8,Fh r theta study q 0,Fh r theta study q -1}.
\item Interestingly in \cref{Fh r theta study q 0.8,Fh r theta study q 0,Fh r theta study q -1}, the sensitivity to $a$ and $\theta$ becomes prominent near the horizon. This was also observed in \cref{Fr r theta study q 0.8,Fr r theta study q 0,F r theta study q -1} for $F_r(r,\theta)$ and in \cref{Ftheta r theta study q 0.8,Ftheta r theta study q 0,Ftheta r theta study q -1} for $F_\theta(r,\theta)$.  
\end{itemize}
Lastly, we study the effect of homogeneous plasma on the shadow of braneworld black hole. As done for profile 1 and profile 2, we plot the shadow of the braneworld black hole for different values of $q, a$ and the plasma parameter $\alpha_3$ in \cref{Plots showing effect of homogeneous plasma  on shadow of braneworld black hole} similar to \cref{Plots showing effect of plasma profile 1 on shadow of braneworld black hole,Plots showing effect of plasma profile 2 on shadow of braneworld black hole}. We observe the following features:
\begin{itemize}
    \item In contrast to inhomogeneous plasma profiles 1 and 2, homogeneous plasma expands the shadow. Particularly, as the plasma parameter $\alpha_3$ approaches 1, the expansion in the shadow size becomes very rapid.
    \item In contrast to profile 1 and profile 2, we continue to get non-circular shadows due to the combined effect of high spin and inclination in presence of homogeneous plasma environment.
    This can be observed, as we go vertically down along each column in \cref{Plots showing effect of homogeneous plasma  on shadow of braneworld black hole}. 
\end{itemize}
Combining observations from \cref{Analysis of plasma profile 1 and its effect on shadow of braneworld black hole,Analysis of plasma profile 2 and its effect on shadow of braneworld black hole,Analysis of homogeneous plasma and its effect on shadow of braneworld black hole}, we find that for all plasma profiles, the shadow size decreases as the tidal charge $q$ increases, with effects illustrated in \cref{Plots showing effect of plasma profile 1 on shadow of braneworld black hole,Plots showing effect of plasma profile 2 on shadow of braneworld black hole,Plots showing effect of homogeneous plasma  on shadow of braneworld black hole}. The deviation from circularity increases with spin and observation angle, but higher  inhomogeneous plasma parameters mitigate this deviation, resulting in nearly circular shadows. Plasma profiles 1 and 2 follow similar trends in shadow size and circularity deviation but profile 2 exerts a higher contraction effect, as detailed in \cref{Plots showing effect of plasma profile 2 on shadow of braneworld black hole}. In contrast, homogeneous plasma uniquely expands the shadow with increasing plasma parameter $\alpha_3$, without affecting circularity deviation due to spin and observation angle, as observed in \cref{Plots showing effect of homogeneous plasma on shadow of braneworld black hole}. These findings highlight the significant role of plasma in shaping braneworld black hole shadow properties, with each profile introducing distinct effects. In what follows, we will investigate the role of the plasma environment on the observed images of M87* and Sgr A*.

\section{EHT Observations of M87* and Sgr A*\label{EHT Observations of M87 and Sgr A}}
We now summarize the Event Horizon Telescope (EHT) observations related to the shadows of M87* and Sgr A*. We also mention the previous estimates of their masses $M$, distances $D$, and inclination angles $\theta_i$.

\paragraph{M87*:}
\begin{enumerate}
   
    \item The EHT collaboration estimated the  vertical angular diameter of the primary ring to be  $\Delta\Theta=(42\pm3)\,\mu\text{as}$  \citep{EventHorizonTelescope:2019dse, EventHorizonTelescope:2019ths, EventHorizonTelescope:2019pgp, EventHorizonTelescope:2019ggy} and reported a maximum offset of $10\%$ between the ring diameter and the shadow diameter \citep{EventHorizonTelescope:2019dse,EventHorizonTelescope:2019ggy}. Considering the   maximum offset of 10\%,  we get  the shadow angular diameter to be  $\Delta\Theta=(37.8\pm3)\,\mu\text{as}$.   
    \item They  also obtained constrains on the deviation from circularity of $\Delta C\lesssim10\%$ and the axis ratio $\Delta A\lesssim 4/3$ \citep{EventHorizonTelescope:2019dse}. Assuming a Kerr black hole, the EHT estimated the mass of M87* to be $M=(6.5\pm0.7)\times10^9M_\odot$ \citep{EventHorizonTelescope:2019dse,EventHorizonTelescope:2019ggy}. 
    \item Previous estimates of mass include $M=6.2^{+1.1}_{-0.6}\times10^9M_\odot$ from stellar dynamics studies \citep{Gebhardt:2009cr,Gebhardt:2000fk} and $M=3.5^{+0.9}_{-0.3}\times10^9M_\odot$ from gas dynamics studies \citep{Walsh:2013uua}. From previous observations the estimated distance and inclination  are  $D=(16.8\pm0.8)\,\text{Mpc}$ \citep{Bird:2010rd} and $\theta_i=(17\pm2)^\circ$ \citep{Tamburini:2019vrf}.
    \item For the mass estimated from stellar dynamics studies and gas dynamics studies the EHT collaboration also provided estimates of Schwarzschild deviation parameter to be $\delta_{sh}=-0.01\pm0.17$\citep{EventHorizonTelescope:2019ggy} and $\delta_{sh}=-0.78\pm0.3$, \citep{EventHorizonTelescope:2019ggy} respectively.  The theoretical $\delta_{sh}$ is defined in \cref{sch deviation parameter}.
\end{enumerate}

\paragraph{Sgr A*:}
\begin{enumerate}
    \item 
The EHT collaboration  reported the primary-ring angular diameter $\Delta\Theta=(51.8\pm2.3)\,\mu\text{as}$ \citep{EventHorizonTelescope:2022wkp, EventHorizonTelescope:2022wok, EventHorizonTelescope:2022xqj, EventHorizonTelescope:2022exc} from the image of Sgr A* and inferred the shadow diameter $\Delta\Theta=(48.7\pm7)\,\mu\text{as}$ \citep{EventHorizonTelescope:2022wkp}. 
\item By studying the motion of the S2 star around Sgr A*, the Keck  team reported $M=(3.975\pm0.058\pm0.026)\times10^6M_\odot$ and $D=(7959\pm59\pm32)\,\text{pc}$ (redshift free), and $M=(3.951\pm0.047)\times10^6M_\odot$, $D=(7935\pm50)\,\text{pc}$ (redshift fixed to unity)  \citep{Do:2019txf}. Independent observations by  the GRAVITY collaboration reported $M=(4.261\pm0.012)\times10^6M_\odot$ and $D=(8246.7\pm9.3)\,\text{pc}$ \citep{GRAVITY:2020gka,GRAVITY:2021xju}; on including optical-aberration systematics, they reported  $M=(4.297\pm0.012\pm0.040)\times10^6M_\odot$ and $D=(8277\pm9\pm33)\,\text{pc}$ \citep{GRAVITY:2020gka,GRAVITY:2021xju}. The estimated inclination angle is $\theta_i=46^\circ$ \citep{Abuter2019A&A...625L..10G}.
\item The EHT collaboration estimated the Schwarzschild deviation parameter as $\delta_{sh}=-0.04^{+0.09}_{-0.1}$ (for the Keck collaboration mass estimate)\citep{EventHorizonTelescope:2022xqj} and$\delta_{sh}=-0.08^{+0.09}_{-0.09}$ (for the GRAVITY collaboration mass estimate)\citep{EventHorizonTelescope:2022wkp,EventHorizonTelescope:2022xqj}. 
\end{enumerate}

\bigskip

In order to obtain constraints on $q$ and $\alpha_i$ with $i=1,2,3$ for profiles 1--3, we use the EHT estimates of the angular diameter  and Schwarzschild deviation parameter $\delta_{sh}$ of the shadow for M87* and Sgr A*.
We follow the procedure as detailed below:

\begin{enumerate}
  \item We select a plasma profile from \cref{profile 1,profile 2,profile 3} and compute the maximum allowed $\alpha_i$ using \cref{alpha max}.
  \item For fixed $\alpha_i$, we vary  $q \in [-2,1]$ and spin $a \in \big[0, \sqrt{1-q}\big]$. The tidal charge cannot have arbitrarily  high negative values, based on different astrophysical observations \citep{Banerjee:2017hzw,Banerjee:2019nnj,Banerjee:2019sae,Bhattacharya:2016naa}. For each $(\alpha_i, q, a)$ combination, we compute the theoretical vertical angular diameter $\Delta\Theta_{\rm th}$ using \cref{theoretical angular diameter} and previously determined mass and distance estimates. We  keep  $\theta_i$ fixed to the central value as the error bars associated with the inclination angle is small and  the inclination angle does not significantly affect the size of the shadow\citep{Sahoo:2023czj}.
 \item We first compare $\Delta\Theta_{th}$ with the EHT estimated diameter $\Delta\Theta_{ EHT}$ and we compute
  $$\chi^2_{\Delta\Theta}=\left(\frac{\Delta\Theta_{ EHT}-\Delta\Theta_{\rm th}(\alpha_i,q,a)}{\sigma_{\Delta\Theta}}\right)^2,$$
  where for M87*, $\Delta\Theta_{ EHT}=37.8\,\mu\text{as}$ and $\sigma=3\,\mu\text{as}$, and for Sgr A* $\Delta\Theta_{ EHT}=48.7\,\mu\text{as}$ and $\sigma=7\,\mu\text{as}$.
  \item For each $(\alpha_i,q)$ we find the spin $a_{\rm min}\in[0,\sqrt{1-q}]$ that minimizes $\chi^2_{\Delta\Theta}$. Thus for a given $(\alpha_i,q)$,  $\chi^2_{\Delta\Theta}(\alpha_i,q)$  is minimum when evaluated at $a_{min}$ as compared to the values obtained for other spin \citep{1976ApAvni}.
  \item We repeat the above procedure for all allowed $\alpha_i$ and $q$, and  then discard combinations of $(\alpha_i,q)$ which give $\chi^2>1$ in allowed range of spin.
  \item We compute $\chi^2_{\delta}$ for the Schwarzschild deviation parameter $\delta_{sh}$ using the estimate provided by the EHT collaboration, 
 $$\chi^2_\delta=\left(\frac{\delta_{sh, EHT }-\delta_{sh,\ th}(\alpha_i,q,a)}{\sigma_{\delta, {EHT}}}\right)^2$$
For  M87*,  $\delta_{sh,\ EHT}=-0.01\pm0.17$ (for stellar dynamics mass estimate), and in case of  Sgr A*,  $\delta_{sh,\ EHT}=-0.04^{+0.09}_{-0.1}$ (for mass and distance measurements estimated by the Keck team)  and $\delta_{sh,\ EHT}=-0.08\pm0.09$ (for mass and distance measurements estimated by the GRAVITY collaboration). 
For the case of Sgr A*, when we use mass and distance estimates reported  by the  Keck collaboration, we take $\sigma_{\delta,\ EHT}=0.1$ for a conservative estimate, although the results would not vary significantly if we take  $\sigma_{\delta,\ EHT}=0.09$. It is important to note that  $\chi^2_{\delta}$  can be expressed in terms of $\chi^2_{\Delta\Theta}$  as,
\begin{gather}
\label{chi delta chi theta relation}
\chi^2_{\delta}=\left(\frac{\sigma_{\Delta\Theta}}{\sigma_{\delta,EHT}\ \Delta\Theta_{Sch}}\right)^2 \chi^2_{\Delta\Theta}
\end{gather}
from above relation it can be inferred that, 
\begin{itemize}
    \item If $\chi^2_{\Delta\Theta}$ is minimum for some $(\alpha,q,a)$ , then so is $\chi^2_{\delta}$.
    \item The parameter space excluded by $\chi^2_{\delta}$ compared to $\chi^2_{\Delta\Theta}$ is more, less or same depending on whether $\left(\frac{\sigma_{\Delta\Theta}}{\sigma_{\delta, EHT}\Delta\Theta_{Sch}}\right)^2$ is greater than 1, less than 1 or equal to 1.
\end{itemize}
  \item We then repeat the entire procedure for the other plasma profiles.
  \item  Finally, we give combined density plots for $\chi^2$ computed for $\Delta\Theta$ and $\delta_{sh}$ as $\chi^2_{\Delta\Theta}$ and $\chi^2_{\delta}$, respectively. The parameter space region common to both $\chi^2_{\Delta\Theta}\leq1$ and $\chi^2_{\delta}\leq1$ gives the allowed range of $\alpha_i$ and $q$. The white region  represents the parameter space excluded outside the $1-\sigma$  of the EHT observations. 
\end{enumerate} 
\subsection{Constraints on tidal charge parameter and plasma environment for M87* \label{Constraints  ontidal charge parameter and plasma environment for M87* considering plasma profile}}
We now proceed to discuss our results for M87* using the methodology described in the previous section.  We assume M87* to be a braneworld black hole surrounded by plasma  profiles 1,2 or 3. We use our methodology to find constraints on the tidal charge $q$ and plasma parameter $\alpha$ using estimates for shadow angular diameter and  Schwarzschild deviation parameter of M87* provided by the EHT collaboration \citep{EventHorizonTelescope:2019ggy}.  \\
When the methodology is applied  considering the mass estimate using gas dynamics studies ($M=3.5\times10^9 M_{\odot}$), it was observed that for the  inhomogeneous plasma profiles 1 and 2,  $\chi^2_{\Delta\Theta}$ and $\chi^2_\delta$ are always greater than unity in the allowed range of $\alpha$, $a$ and $-2\leq q\leq1$. Thus, for the cases of profiles 1 and 2 we provide density plots for mass estimate from stellar dynamics studies ($M=6.2\times10^9 M_{\odot}$) and the EHT collaboration ($M=6.5\times10^9M_{\odot}$). The analysis and density plots considering mass estimate reported by the EHT collaboration are for the sake of completeness, because the EHT collaboration already presumes M87*  to be a Kerr black hole when estimating its mass from the shadow.  
\subsubsection{Constraints on tidal charge considering inhomogeneous plasma profile 1\label{Constraints from density plots  for M87* considering plasma profile 1}}
\begin{figure}[h!]
    \centering
    \begin{subfigure}[b]{0.45\textwidth}
        \centering
        \includegraphics[width=\textwidth]{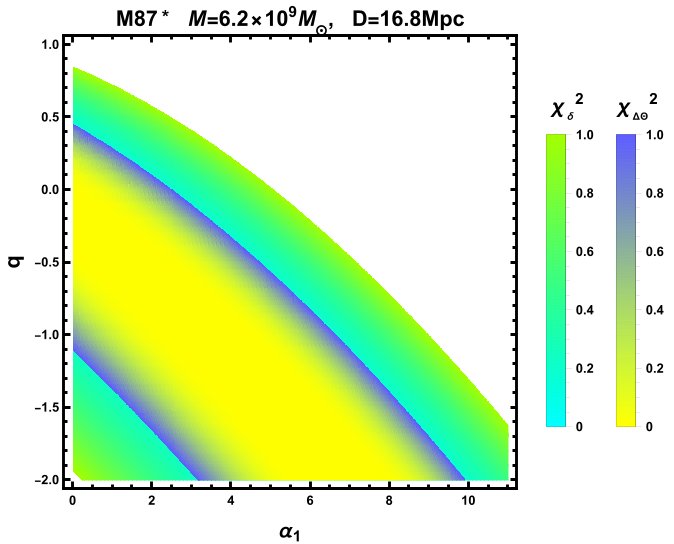} 
        \caption{ \label{r_profile_combined_density_plot_M87_6.2.jpg}}
    \end{subfigure}
    \hfill
    \begin{subfigure}[b]{0.45\textwidth}
        \centering
        \includegraphics[scale=0.56]{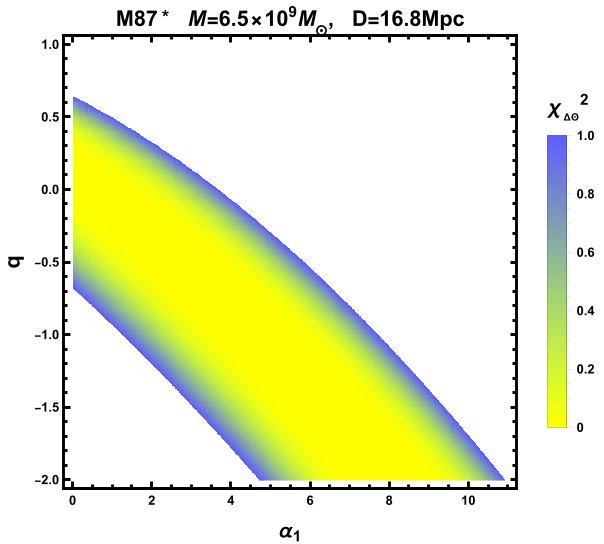} 
        \caption{ \label{r_profile_angular_diameter_density_plot_M87_6.5.jpg}}
      \end{subfigure}

   \caption{\Cref{r_profile_combined_density_plot_M87_6.2.jpg,r_profile_angular_diameter_density_plot_M87_6.5.jpg}  respectively show the combined density plot (for $\chi^2_{\Delta\Theta}$ and $\chi^2_{\delta}$)  and $\chi^2_{\Delta\Theta}$  considering plasma profile 1  using mass estimate from (a) stellar dynamics studies and (b) the EHT collaboration for M87*. The inclination angle $\theta_i=17^{\circ}$. \label{m87_density_plot profile1}}
\end{figure}

     In \cref{m87_density_plot profile1} we show the variation of $\chi^2$ with $q$ and plasma parameter $\alpha_1$ within error bars of the  EHT estimates for M87* . \cref{r_profile_combined_density_plot_M87_6.2.jpg}
represents combined density plot showing the variation of $\chi^2_{\Delta\Theta}$ and $\chi^2_\delta$  for mass estimate of M87* considering stellar dynamics studies.  
 For the case of mass estimated by the  EHT collaboration, we have only plotted density plots for $\chi^2 _{\Delta\Theta}$ in \cref{{r_profile_angular_diameter_density_plot_M87_6.5.jpg}} as the estimate for Schwarzschild deviation parameter $\delta_{sh}$ is not provided by the EHT team.  For this reason, we have also done the same for the cases when we consider profile 2 and profile 3 (homogeneous plasma). 
We now make the following observations from \cref{m87_density_plot profile1}:
\begin{itemize}
    \item In \cref{r_profile_combined_density_plot_M87_6.2.jpg} , it can be observed that the parameter space  constrained by angular diameter through $\chi^2_{\Delta\Theta}$ is more stringent compared to Schwarzschild deviation parameter $\delta_{sh}$. This is because, for the case of M87* considering mass estimate from stellar dynamics studies $\sigma_{\Delta\Theta}<(\sigma_{\delta, EHT}\  \Delta\Theta_{Sch})$ (refer \cref{chi delta chi theta relation}), hence  $\chi^2_{\delta}$ can accommodate larger parameter space.
    This is also observed in density plots of   \cref{m87_density_plot profile2,m87_density_plot profile3}.
    \item When we consider  $\alpha_1$ to be non-zero, the allowed range of tidal charge within $1-\sigma$ varies with $\alpha_1$.  In both \cref{r_profile_combined_density_plot_M87_6.2.jpg,r_profile_angular_diameter_density_plot_M87_6.5.jpg} it can be observed that as $\alpha_1$ increases we can accommodate more negative tidal charge to reproduce the data.  
    \item In the absence of plasma ($\alpha_1 \approx 0$), from $\chi^2_{\Delta\Theta}$ we rule out $q\gtrsim0.45$ and $q\lesssim-1.15$ outside $1-\sigma$  . Thus, the allowed range of tidal charge  is $-1.15\lesssim q\lesssim0.45$, which includes the Kerr scenario but does not rule out the non-zero tidal charge scenario; particularly the negative tidal charge scenario  which a confirmatory signature of extra dimensions.  
    We consider $\chi^2_{\Delta\Theta}$ to establish this range in $q$ as it offers a more stringent constraint.
    \item In the absence of plasma, when we consider the  density plot computed  using mass estimate by the EHT collaboration, we observe that the allowed range of tidal charge is $-0.7\lesssim q\lesssim0.65$. We note that the lower bound on the tidal charge is less in \cref{r_profile_combined_density_plot_M87_6.2.jpg} than in \cref{r_profile_angular_diameter_density_plot_M87_6.5.jpg}, which may be attributed to the higher mass reported by the EHT compared to the mass estimated from stellar dynamics studies.

\end{itemize}
\subsubsection{Constraints on tidal charge considering inhomogeneous plasma profile 2 \label{Constraints from density plots  for M87* considering plasma profile 2}}

\begin{figure}[h!]
    \centering
    \begin{subfigure}[b]{0.45\textwidth}
        \centering
        \includegraphics[width=\textwidth]{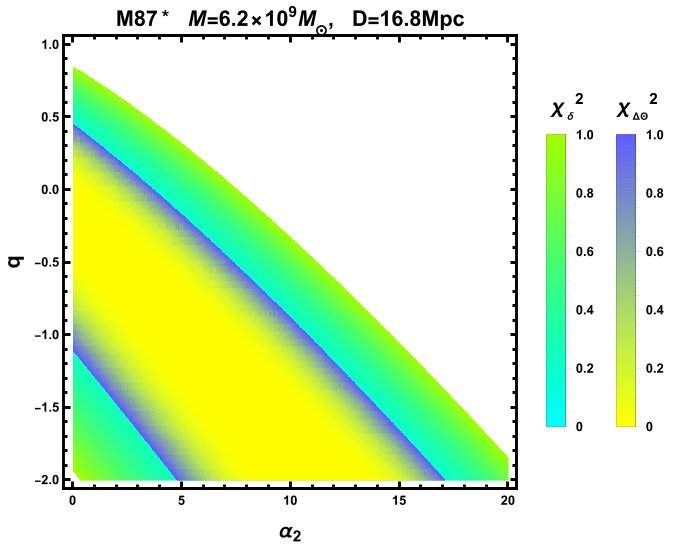} 
        \caption{ \label{theta_profile_combined_density_plot_M87_6.2.jpg}}
    \end{subfigure}
    \hfill
    \begin{subfigure}[b]{0.45\textwidth}
        \centering
        \includegraphics[width=\textwidth]{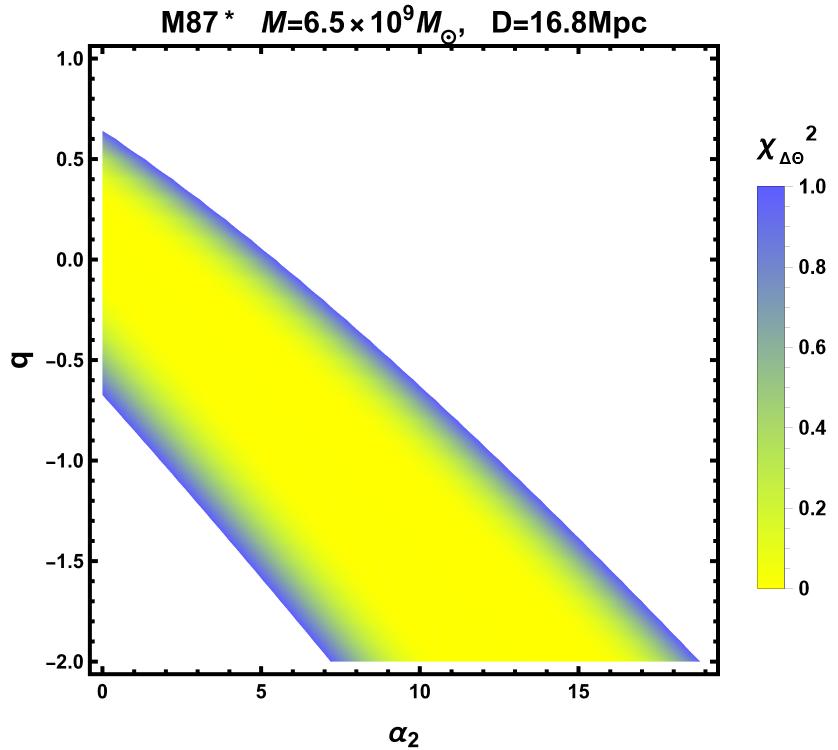} 
        \caption{ \label{theta_profile_angular_diameter_density_plot_M87_6.5.jpg}}
      \end{subfigure}

   \caption{\Cref{theta_profile_combined_density_plot_M87_6.2.jpg,theta_profile_angular_diameter_density_plot_M87_6.5.jpg}  show the combined density plot (for $\chi^2_{\Delta\Theta}$ and $\chi^2_{\delta}$)  and $\chi^2_{\Delta\Theta}$  considering plasma profile 2  using mass estimate from stellar dynamics studies and the EHT collaboration for M87*, respectively. The inclination angle $\theta_i=17^{\circ}$. \label{m87_density_plot profile2}}

\end{figure}

     We now discuss constraints on the tidal charge $q$  and plasma parameter $\alpha_2$ from density plots  in \cref{m87_density_plot profile2}. We observe the following features:
\begin{itemize}
    \item In case of profile 2, from \cref{theta_profile_combined_density_plot_M87_6.2.jpg} we can observe that the constraints on parameter space are more stringent due to $\Delta\Theta$ compared to $\delta_{sh}$.
     \item Similar to the case of plasma profile 1, we observe that as $\alpha_2$ increases the upper and lower bounds on the allowed tidal charge $q$  decrease. 
    \item    In the absence of plasma ($\alpha_2=0$), from density plots for the mass estimates from stellar dynamics studies    (see \cref{theta_profile_combined_density_plot_M87_6.2.jpg}), the allowed range of tidal charge within $1-\sigma$ is $-1.15\lesssim q\lesssim0.45$. In the case of density plots for mass estimated by the EHT  collaboration, the allowed range for tidal charge is $-0.7\lesssim q\lesssim0.65$ as  seen in \cref{theta_profile_angular_diameter_density_plot_M87_6.5.jpg}. In the absence of plasma we are simply dealing with the brane world black hole, thus, it is expected that the constraints on tidal charge will be the same irrespective of the plasma profile we consider.  
   
    \item For $q\simeq-2$, the lower and upper bound on $\alpha_2$ allowed within $1-\sigma $  are 5 and 17 for mass estimated from stellar dynamics  studies (refer \cref{theta_profile_combined_density_plot_M87_6.2.jpg}), respectively. These are higher than the upper and lower bounds for $q\simeq-2$ found in case of profile 1 (refer \cref{r_profile_combined_density_plot_M87_6.2.jpg}) which are $\alpha_1=10$ and $3$, respectively. This indicates that for a given $q$, profile 2 is able to cause same contraction as profile 1 with a higher plasma  parameter. Thus, for the case of M87* we see that profile 1 has a greater contracting effect compared to profile 2.
It is important to note that the relative contracting effect of profile 1 and profile 2 depends on the inclination angle of the observer, e.g. in \cref{Analysis of plasma profile 2 and its effect  on shadow of braneworld black hole} where the inclination angle was taken to be $90^\circ$ we saw that the profile 2 contracts the shadow more compared to profile 1.
 
    \item It can be concluded from \cref{m87_density_plot profile1,m87_density_plot profile2} that, in order to reproduce the data high density plasma favours the negative tidal charge scenario.
\end{itemize}


\subsubsection{Constraints on tidal charge considering homogeneous plasma profile\label{Constraints from density plots  for M87* considering plasma profile 3}}

\begin{figure}[h!]
    \centering
    \begin{subfigure}[b]{0.45\textwidth}
        \centering
        \includegraphics[width=\textwidth]{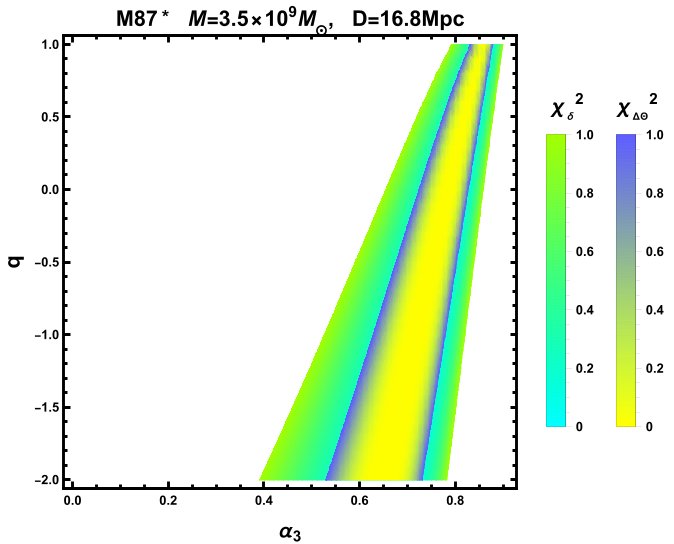} 
        \caption{ \label{homogeneous_profile_combined_density_plot_M87_3.5.jpg}}
        
    \end{subfigure}
    \hfill
    \begin{subfigure}[b]{0.45\textwidth}
        \centering
        \includegraphics[width=\textwidth]{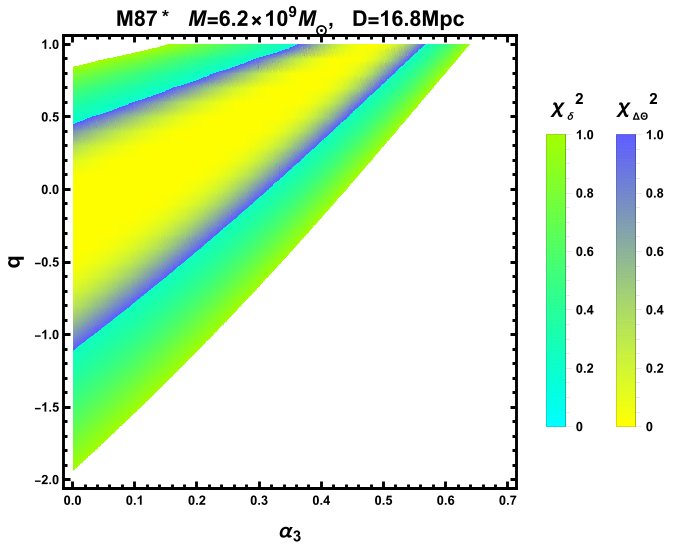} 
        \caption{ \label{homogeneous_profile_combined_density_plot_M87_6.2.jpg}}
            \end{subfigure}
\vspace{0.5cm}

    \begin{subfigure}[b]{0.45\textwidth}
        \centering
        \includegraphics[width=\textwidth]{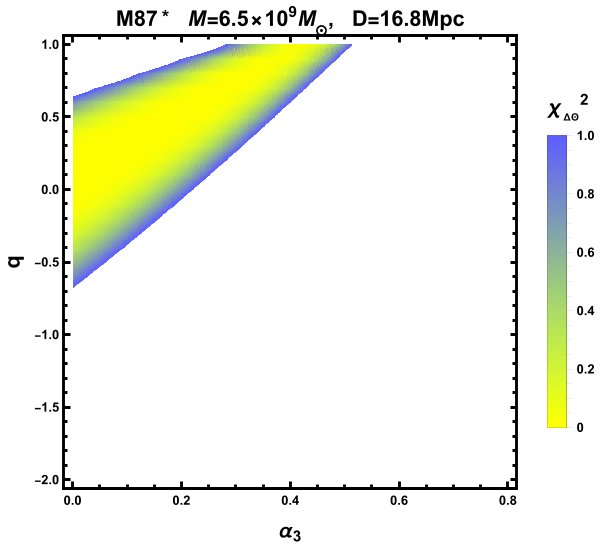} 
        \caption{ \label{homogeneous_profile_angular_diameter_density_plot_M87_6.5.jpg}}
        
    \end{subfigure}
   \caption{\Cref{homogeneous_profile_combined_density_plot_M87_3.5.jpg,homogeneous_profile_combined_density_plot_M87_6.2.jpg,homogeneous_profile_angular_diameter_density_plot_M87_6.5.jpg}  show the combined density plot (for $\chi^2_{\Delta\Theta}$ and $\chi^2_{\delta}$)  and $\chi^2_{\Delta\Theta}$  considering homogeneous plasma  using mass estimates from gas dynamics studies, stellar dynamics studies and the EHT collaboration for M87*, respectively. The inclination angle $\theta_i=17^{\circ}$. \label{m87_density_plot profile3}}

\end{figure}

     For the sake of completeness, we also discuss density plots considering homogeneous plasma around M87*.   In \cref{m87_density_plot profile3}, \cref{homogeneous_profile_combined_density_plot_M87_3.5.jpg,homogeneous_profile_combined_density_plot_M87_6.2.jpg} are combined density plots for mass estimates from gas dynamics and stellar dynamics studies, respectively. \cref{homogeneous_profile_angular_diameter_density_plot_M87_6.5.jpg}  represents the angular diameter density plots for mass of M87* estimated by the EHT collaboration. For reasons discussed  in  \cref{Constraints from density plots  for M87* considering plasma profile 1}, we observe that in \cref{m87_density_plot profile3} parameter space constraints due to $\Delta\Theta$ is more stringent compared to $\delta_{sh}$.   We highlight  the following distinct features from the density plots:
    \begin{itemize}
        \item Firstly, unlike in the case of inhomogeneous plasma (profiles 1 and 2) we find that,  there exists a parameter space allowed within $1-\sigma$  of the EHT observations when the mass estimate of M87* from gas dynamics studies is considered. This can be observed from \cref{homogeneous_profile_combined_density_plot_M87_3.5.jpg}. The primary reason behind this is,  the expansive effect of homogeneous plasma parameter which is able to compensate for the small mass estimate from gas dynamics studies ($M=3.5\times10^9 M_\odot$),  used to compute the theoretical angular diameter and the Schwarzschild deviation parameter.
        \item Furthermore in \cref{homogeneous_profile_combined_density_plot_M87_3.5.jpg}, we  observe that there is a lower bound on plasma parameter $\alpha_3\sim0.52$ , below which no value of tidal charge in the range $-2\leq q\leq1$ can explain $\Delta\Theta$ and $\delta_{sh}$ within $1-\sigma$ of the EHT estimates. However, for plasma parameter $\alpha_3\gtrsim0.52$  we observe that  for every $q$ in the range $-2\leq q\leq1$, there exists an upper and lower bound of $\alpha_3$ which can reproduce both $\Delta\Theta$ and $\delta_{sh}$  within  $1-\sigma$ error bars estimated by the EHT collaboration.
        \item In the absence of plasma ($\alpha_3=0$), for the case of mass estimates of M87* reported from stellar dynamics studies and the EHT collaboration, the allowed range of tidal charge are the same as reported before in \cref{Constraints from density plots  for M87* considering plasma profile 1,Constraints from density plots  for M87* considering plasma profile 2}. For reasons discussed in previous points, the no plasma scenario is completely ruled out when mass estimate from gas dynamics studies is considered (refer \cref{homogeneous_profile_combined_density_plot_M87_3.5.jpg}).       
        \item However, comparing \cref{m87_density_plot profile1,m87_density_plot profile2,m87_density_plot profile3},  we observe that in contrast to the cases of profiles 1 and 2, an increase in plasma parameter $\alpha_3$ favours a positive tidal charge scenario. This is primarily because of the expanding effect caused by the homogeneous plasma, such that, in order to reproduce the observed $\Delta\Theta$ or $\delta_{sh}$ at high plasma parameter, the contracting effect of positive tidal charge $q$ is required.  
\end{itemize}

  From the analysis of density plots from \cref{m87_density_plot profile1,m87_density_plot profile2,m87_density_plot profile3} we observe that when plasma is considered, the constraints on tidal charge can vary significantly as plasma density is varied (in terms of the plasma parameter).  Secondly,  the available observational constraints from  $\Delta\Theta$ and $\delta_{sh}$ (in the case of M87*) provide further restrictions on upper bounds on the plasma parameter compared to the theoretical bound for a given tidal charge $q$ (refer \cref{Analysis of plasma profile 1 and its effect  on shadow of braneworld black hole,Analysis of plasma profile 2 and its effect  on shadow of braneworld black hole,Analysis of homogeneous plasma  and its effect  on shadow of braneworld black hole}).     Additionally, if we take into account  the electron number density estimate $n_e$ near M87* then we can give more stringent constraints on the parameter space. 
  The plasma parameter $\alpha_i$ can be estimated from  the electron number density $n_e$  considering the relation obtained using \cref{plasma frequency electron number density relation,scaled plasma frequency},
\begin{gather}
    \label{alpha and ne relation }
    \alpha_i=\frac{4\pi e^2}{m_e\omega^2_0} \, \frac{\rho^2 n_e(r,\theta)}{f(r)+g(\theta)}
\end{gather}
In the above equation $i=1,2,3$ represents the plasma parameters for profiles 1, 2, and 3 as described in \cref{profile 1,profile 2,profile 3}, respectively. Using $f(r)$ and $g(\theta)$ definitions for profiles 1,2 and 3 from \cref{profile 1,profile 2,profile 3} we obtain:
\begin{gather}
    \label{alpha1 and ne relation}
    \alpha_1=\frac{4\pi e^2}{m_e\omega^2_0} \, \frac{\rho^2\ n_e(r,\theta)}{\sqrt{r}}\\
    \label{alpha2 and ne relation}
    \alpha_2=\frac{4\pi e^2}{m_e\omega^2_0} \, \frac{\rho^2\ n_e(r,\theta)}{1+2 \sin^2\theta}\\
    \label{alpha3 and ne relation}
    \alpha_3=\frac{4\pi e^2}{m_e\omega^2_0} \, n_e(r,\theta)\\
    \text{where, $\rho^2=r^2+a^2\cos^2\theta$}\nonumber
\end{gather}
Above equations can be used to estimate plasma parameter $\alpha_i$ if estimates of $n_e(r,\theta)$ at a given location are known,  {along with  the asymptotic frequency of the  photon $\omega_0$ } .  It is interesting to note that since $\rho^2,f(r)$ and $g(\theta)$ do not explicitly depend on tidal charge $q$, therefore the estimates of $\alpha_i$ will be the same as that of assuming Kerr scenario, irrespective of the tidal charge $q$.
The EHT collaboration has reported the electron number density $n_e\sim10^4-10^7$ cm$^{-3}$  at $r\sim5 r_g$ and $\theta=\pi/2$ for M87* assuming a one-zone isothermal sphere model \citep{EventHorizonTelescope:2021srq}.  {We set  $\omega_0\sim230$  GHz (the frequency of observation of the EHT) for computing the plasma parameters.} We provide upper and lower bounds of $\alpha_1,\alpha_2$ and $\alpha_3$ using \cref{alpha1 and ne relation,alpha2 and ne relation,alpha3 and ne relation} :

\begin{table}[h!]
    \centering
    \begin{tabular}{|c|c|c|c|}
        \hline
         $n_e\sim$ \citep{EventHorizonTelescope:2021srq} & $\alpha_1\sim$  & $\alpha_2\sim$  & $\alpha_3\sim$ \\
         \hline
      $10^4$ cm$^{-3}$   & $1.7\times 10^{-10}$ &$3.81\times 10^{-10}$  & $1.52\times 10^{-11}$ \\
         \hline
         $10^7$ cm$^{-3}$ & $1.7\times 10^{-7}$ &$3.81\times 10^{-7}$  & $1.52\times 10^{-8}$ \\
    \hline
    \end{tabular}

    \caption{Estimates for $\alpha_1$, $\alpha_2$ and $\alpha_3$ corresponding to lower and upper bounds of $n_e$ for M87* computed at $r=5 r_g$ and $\theta=\pi/2$ using \citep{EventHorizonTelescope:2021srq} . \label{table for alpha estimate from ne}}
   
\end{table}
For the case of profile 1 using  \cref{profile 1}, it is possible to relate the plasma parameter $\alpha_1$ with  the mass accretion rate $\dot{M}$, assuming neutral hydrogen cold plasma as described by \citep{Perlick:2015vta}:
\begin{gather}
    \label{perlick mdot alpha}
    \alpha_1=\frac{e^2\ \dot{M}\ c^3}{\sqrt{2} m_e m_p \omega^2_0 G^2 M^2}
\end{gather}
 We use  \cref{perlick mdot alpha}     to estimate $\alpha_1$ from $\dot{M}$. The EHT collaboration reports an accretion rate $\dot{M}\sim (3-20)\times 10^{-4} M_\odot $ yr$^{-1}$ \citep{EventHorizonTelescope:2021srq}  while Drew et.al \citep{Drew:2025euq} reported the accretion rate to be $4\times10^{-5}M_{\odot}$ yr$^{-1} -0.4{M_{\odot}}$ yr$^{-1}$.  Considering the upper bounds on the accretion rate from  both the estimates we report below the maximum magnitude of $\alpha_1$
\begin{table}[h!]
    \centering
    \begin{tabular}{|c|c|}
    \hline
       maximum $\dot{M}\sim$ & $\alpha_1\sim$ \\
       \hline
         $2\times 10^{-3} M_\odot$ yr$^{-1}$ \citep{EventHorizonTelescope:2021srq}&  $2.58\times10^{-10}$\\
         \hline
         $0.4 M_\odot$ yr$^{-1}$ \citep{Drew:2025euq} & $5.15 \times 10^{-8}$ \\
         \hline
    \end{tabular}
    \caption{Estimates of $\alpha_1$ based  on maximum estimate of accretion rates 
    . \label{alpha1 estimate from mdot}}
    
\end{table}

From \cref{table for alpha estimate from ne,alpha1 estimate from mdot}, it is reasonable to expect constraints for $\alpha_i\sim0$  as most relevant. Thus, from \cref{homogeneous_profile_combined_density_plot_M87_3.5.jpg} if $\alpha_3\sim0$ is considered, then neither the Kerr  nor the braneworld scenario are able to explain the EHT estimates of $\Delta \Theta$ and $\delta_{sh}$, when mass estimate from gas dynamics is considered.  Therefore, the angular diameter, the Schwarzschild deviation parameter and the electron number density, when combined, can give strong restrictions on the modified theory of gravity.   

\subsection{Constraints on tidal charge parameter and plasma environment for Sgr A* \label{Constraints  tidal charge parameter and plasma environment for Sgr A* considering plasma profile} }
We now discuss the constraints on the tidal charge parameter $q$ and the plasma parameter $\alpha$  from the EHT estimates of shadow angular diameter $\Delta\Theta_{obs}$ and Schwarzschild deviation parameter $\delta_{sh}$   for Sgr A* considering  plasma profiles 1, 2 and 3. For the mass and distance estimated by the Keck team and the GRAVITY collaboration, the EHT collaboration reported $\delta_{sh}=-0.04^{+0.09}_{-0.1}$ and $\delta_{sh}=-0.08^{+0.09}_{-0.09}$\citep{EventHorizonTelescope:2022xqj}, respectively.  The angular diameter of the shadow, inferred from the primary ring for Sgr A* by the EHT collaboration is  $\Delta\Theta_{obs}=(48.7\pm7)\mu as$\citep{EventHorizonTelescope:2022xqj}. 
For the parameter space constrained by $1-\sigma$ interval of $\Delta\Theta_{obs}$ and $\delta_{sh}$, the variation of $\chi^2_{\Delta\Theta}$ and $\chi^2_{\delta}$ are shown in the form of combined density plots with tidal charge $q$ along $y$-axis and plasma parameter $\alpha$ along $x$-axis in \cref{sgra_density_plots_profile1,sgra_density_plots_profile2,sgra_density_plots_profile3}.    
The combined density plots for $\Delta\Theta$ and $\delta_{sh}$  highlight the common  allowed parameter space.
In the case of Sgr A*, $\sigma_{\delta, EHT}\Delta\Theta_{Sch}<\sigma_{\Delta\Theta}$. Therefore, unlike the case of M87* (refer \cref{Constraints  ontidal charge parameter and plasma environment for M87* considering plasma profile} ) we find that in combined density plots for plasma profiles 1,2 and 3, $\delta_{sh}$ gives more stringent constraints compared to $\Delta\Theta$ as seen in \cref{sgra_density_plots_profile1,sgra_density_plots_profile2,sgra_density_plots_profile3}. 
\subsubsection{Constraints on tidal charge considering inhomogeneous plasma profile 1\label{Constraints from density plots  for Sgr A* considering plasma profile 1}}

\begin{figure}[h!]
    \centering
    \begin{subfigure}[b]{0.45\textwidth}
        \centering
        \includegraphics[width=\textwidth]{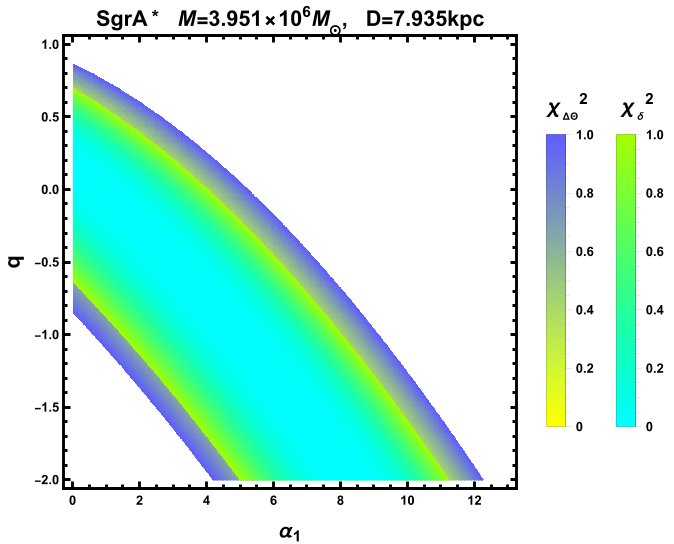} 
        \caption{Keck Team \label{r_profile_combined_density_plot_SgrA_3.951.jpg}}
    \end{subfigure}
    \hfill
    \begin{subfigure}[b]{0.45\textwidth}
        \centering
        \includegraphics[width=\textwidth]{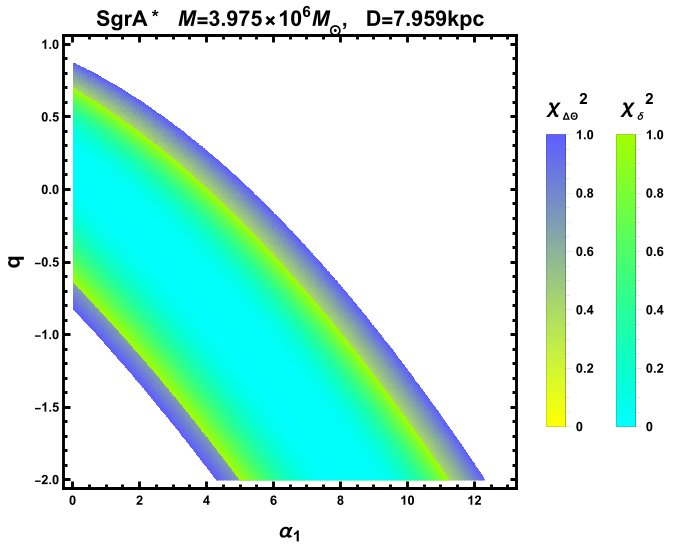} 
        \caption{Keck Team \label{r_profile_combined_density_plot_SgrA_3.975.jpg}}
      \end{subfigure}
\vspace{0.5cm}
\begin{subfigure}[b]{0.45\textwidth}
        \centering
        \includegraphics[width=\textwidth]{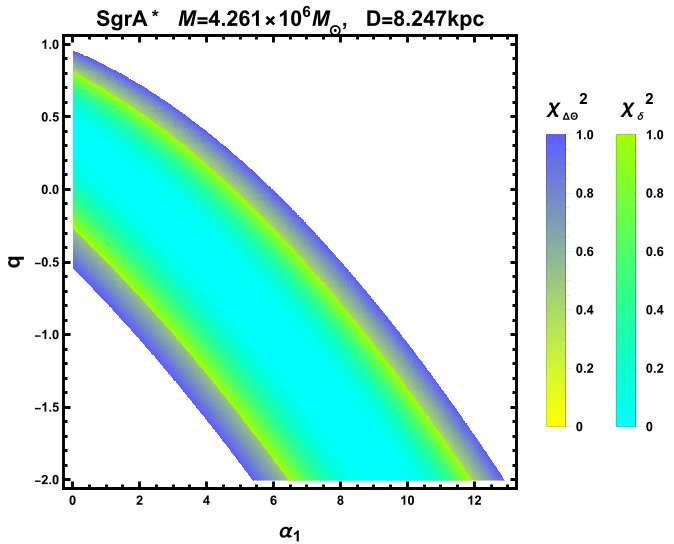} 
        \caption{ GRAVITY collaboration\label{r_profile_combined_density_plot_SgrA_4.261.jpg}}
        
    \end{subfigure}
    \hfill
    \begin{subfigure}[b]{0.45\textwidth}
        \centering
        \includegraphics[width=\textwidth]{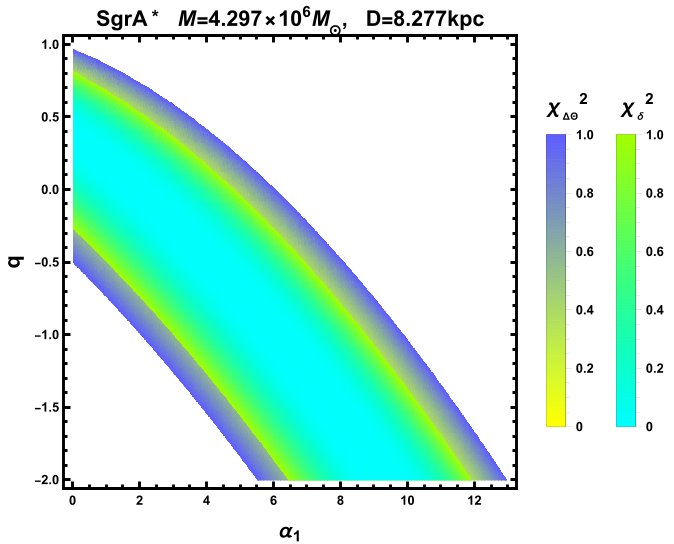} 
        
        \caption{GRAVITY collaboration \label{r_profile_combined_density_plot_SgrA_4.297}}
    \end{subfigure}
   \caption{The  \cref{r_profile_combined_density_plot_SgrA_3.951.jpg,r_profile_combined_density_plot_SgrA_3.975.jpg,r_profile_combined_density_plot_SgrA_4.261.jpg,r_profile_combined_density_plot_SgrA_4.297} represent combined density plots of the parameter space constrained by the EHT estimate of shadow angular diameter and  Schwarzschild deviation parameter  for both the mass and distance estimates by the Keck  team and  the GRAVITY collaboration, respectively.  All the plots are computed considering plasma profile 1 and taking $\theta_i=46^\circ$. The overlapping region represents the allowed parameter space from  both the observables, $\Delta\Theta_{obs}$  and $\delta_{sh}$.  \label{sgra_density_plots_profile1}}
\end{figure}

We first discuss our results for Sgr A* considering   plasma profile 1. The combined density plots are shown in \cref{sgra_density_plots_profile1}.  \cref{r_profile_combined_density_plot_SgrA_3.951.jpg,r_profile_combined_density_plot_SgrA_3.975.jpg} show the variation of $\chi^2_{\Delta\Theta}$ and $\chi^2_{\delta}$ considering mass and distance estimates by the Keck team, keeping the redshift parameter free and then fixing it to unity, respectively. Similarly, \cref{r_profile_combined_density_plot_SgrA_4.261.jpg,r_profile_combined_density_plot_SgrA_4.297}   show the variation of $\chi^2_{\Delta\Theta}$ and $\chi^2_{\delta}$ considering mass and distance estimates by the GRAVITY collaboration, without considering effects of optical aberrations  and then improving their estimates by taking into account optical aberrations, respectively.  The following can be observed from the density plots:
\begin{itemize}
    \item  The parameter space constrained by $\chi^2_{\delta}$ is more stringent compared to the parameter space constrained by $\chi^2_{\Delta\Theta}$. This can be observed in \cref{r_profile_combined_density_plot_SgrA_3.951.jpg,r_profile_combined_density_plot_SgrA_3.975.jpg,r_profile_combined_density_plot_SgrA_4.261.jpg,r_profile_combined_density_plot_SgrA_4.297}. 
    \item  The density plots obtained by using mass and distance reported by the Keck team are shown in \cref{r_profile_combined_density_plot_SgrA_3.951.jpg,r_profile_combined_density_plot_SgrA_3.975.jpg}. From the plots we note that when  $\alpha_1\simeq 0$, $q\gtrsim0.7$ and $q\lesssim-0.65$ are ruled out outside $1-\sigma$ from the joint constraints based on $\Delta\Theta_{obs}$ and $\delta_{sh}$.     
    \item In  \cref{r_profile_combined_density_plot_SgrA_4.261.jpg,r_profile_combined_density_plot_SgrA_4.297},  the density plots obtained by using mass and distance estimates reported by the GRAVITY collaboration are shown. From the plots we observe that in the absence of plasma ($\alpha_1\simeq 0$), the allowed range of tidal charge corresponds to $-0.3\lesssim q\lesssim0.8$ when constrained within $1-\sigma$ of  $\delta_{sh}$.  
    \item Furthermore, in the presence of plasma, when mass and distance  estimates by the Keck team and the GRAVITY collaboration are considered, from  \cref{sgra_density_plots_profile1} we observe that $\alpha_1\gtrsim11$ and $\alpha_1\gtrsim12$ are respectively ruled out outside of $1-\sigma$ for $q\simeq -2$. Note that, different astrophysical observations prevent $q$ to assume arbitrarily high negative values \citep{Banerjee:2017hzw,Banerjee:2019nnj,Banerjee:2019sae,Bhattacharya:2016naa}. Hence, the shadow related data of Sgr A* prevent $\alpha_1$ to be greater than 12.

    \item As was observed in the case of M87* with plasma profile 1 (in \cref{Constraints from density plots  for M87* considering plasma profile 1}), the upper and lower bounds of tidal charge $q$ for the common parameter space constrained by $\chi^2_{\delta}\leq1$ and $\chi^2_{\Delta\Theta}\leq1$ decrease as $\alpha_1$ increases. 
\end{itemize}

 
\subsubsection{Constraints on tidal charge considering inhomogeneous plasma profile 2\label{Constraints from density plots  for Sgr A* considering plasma profile 2}}
\begin{figure}[h!]
    \centering
    \begin{subfigure}[b]{0.45\textwidth}
        \centering
        \includegraphics[width=\textwidth]{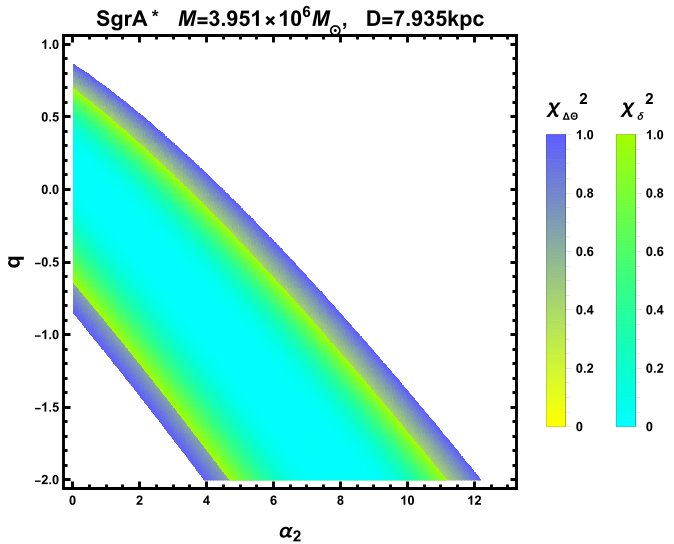} 
        \caption{ Keck team \label{theta_profile_combined_density_plot_SgrA_3.951.jpg}}
        
    \end{subfigure}
    \hfill
    \begin{subfigure}[b]{0.45\textwidth}
        \centering
        \includegraphics[width=\textwidth]{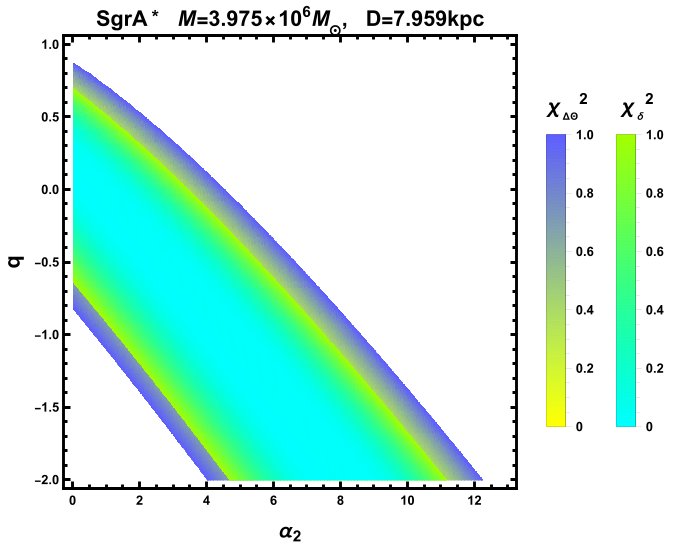} 
        \caption{Keck team  \label{theta_profile_combined_density_plot_SgrA_3.975.jpg} }
       
    \end{subfigure}
\vspace{0.5cm}
\begin{subfigure}[b]{0.45\textwidth}
        \centering
        \includegraphics[width=\textwidth]{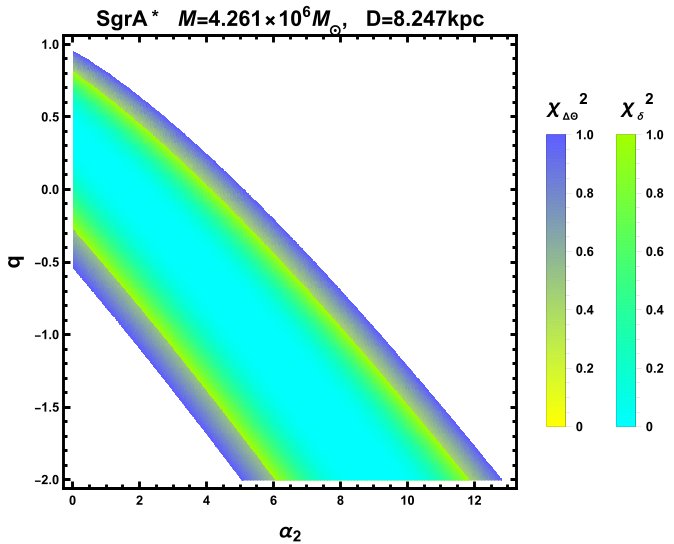} 
        \caption{GRAVITY collaboration \label{theta_profile_combined_density_plot_SgrA_4.261.jpg} }
       
    \end{subfigure}
    \hfill
    \begin{subfigure}[b]{0.45\textwidth}
        \centering
        \includegraphics[width=\textwidth]{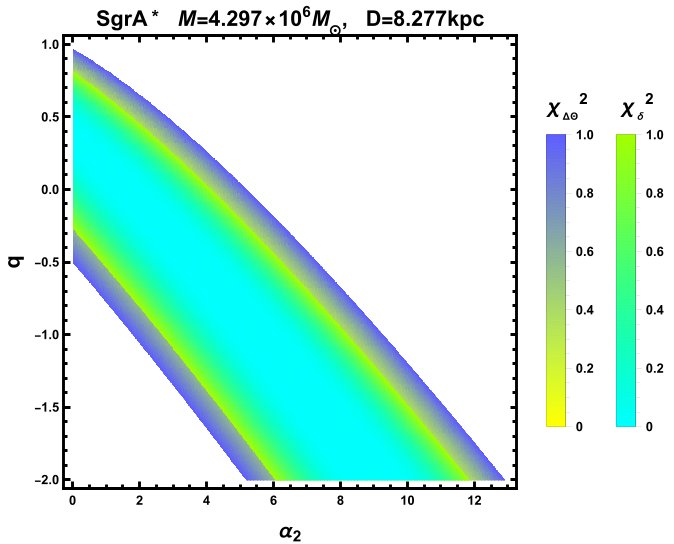} 
        \caption{GRAVITY collaboration \label{theta_profile_combined_density_plot_SgrA_4.297} }
        
    \end{subfigure}
   \caption{The  \cref{theta_profile_combined_density_plot_SgrA_3.951.jpg,theta_profile_combined_density_plot_SgrA_3.975.jpg,theta_profile_combined_density_plot_SgrA_4.261.jpg,theta_profile_combined_density_plot_SgrA_4.297} represent combined density plots of the allowed parameter space constrained by the EHT estimate of the shadow angular diameter and the Schwarzschild deviation parameter  for both the mass and distance estimates by the Keck team and the GRAVITY collaboration, respectively.  All the plots are computed considering plasma profile 2 and taking $\theta_i=46^\circ$. The overlapping region represents the allowed parameter space by both the observables, $\Delta\Theta_{obs}$  and $\delta_{sh}$.  \label{sgra_density_plots_profile2}}
\end{figure}

We now proceed to discuss combined density plots in \cref{sgra_density_plots_profile2} for Sgr A* considering plasma profile 2. \cref{theta_profile_combined_density_plot_SgrA_3.951.jpg,theta_profile_combined_density_plot_SgrA_3.975.jpg} represent combined density plots considering the Keck team mass and distance estimates. The combined density plots considering mass and distance estimates reported by the GRAVITY collaboration  are shown in \cref{theta_profile_combined_density_plot_SgrA_4.261.jpg,theta_profile_combined_density_plot_SgrA_4.297}. We highlight the following features from the analysis of the density plots:
\begin{itemize}
    \item In the absence of plasma ($\alpha_2\simeq 0$), the constraints on tidal charge are same as that observed in case of plasma profile 1 (refer  \cref{Constraints from density plots  for Sgr A* considering plasma profile 1}), i.e., $-0.65\lesssim q\lesssim0.7$ from the Keck team mass estimates and $-0.3\lesssim q\lesssim0.8$ from the GRAVITY collaboration mass estimates.
   
    \item In the presence of plasma profile 2, we observe that the allowed lower and upper bounds of $q$ decrease as $\alpha_2$ increases. The nature of variation of the bounds on $q$ with $\alpha_2$ for Sgr A* is similar to the case of M87*. This can be observed by comparing \cref{m87_density_plot profile2,sgra_density_plots_profile2}.
    \item Furthermore, from \cref{sgra_density_plots_profile2}, we rule out $\alpha_2\gtrsim11$ for the case of mass and distance estimates by the Keck team and $\alpha_2\gtrsim12$ for the case of mass and distance estimates by the GRAVITY collaboration, when $q\simeq -2$. The bounds on the plasma parameter are subject to increase as $q$ becomes more and more negative. However, based on different astrophysical observations the tidal charge cannot have arbitrarily  high negative values \citep{Banerjee:2017hzw,Banerjee:2019nnj,Banerjee:2019sae,Bhattacharya:2016naa}.

 \item It is interesting to note that unlike M87* (refer \cref{m87_density_plot profile1,m87_density_plot profile2}), the allowed upper bounds on 
 the two inhomogeneous plasma parameters $\alpha_{1,max}$ and $\alpha_{2,max}$ (when $q\simeq -2$) are nearly the same (refer \cref{sgra_density_plots_profile1,sgra_density_plots_profile2})
in case of Sgr A*. However, when $q\sim 0$ is considered, then $\alpha_{2,max}< \alpha_{1,max}$ for the case of Sgr A*. However for M87*, irrespective of whether $q\sim 0$ or negative (e.g. $q\simeq -2$), $\alpha_{1,max}< \alpha_{2,max}$. This implies that the contracting effect of the inhomogeneous plasma profile is sensitive to the distance-to-mass ratio of the source and the inclination angle, such that for M87*, plasma profile 1 has a greater contracting effect compared to plasma profile 2, while the opposite holds for Sgr A*. We further note that for Sgr A*, as $q$  becomes more and more negative, the difference between $\alpha_{2,max}$ and $\alpha_{1,max}$ decreases, which indicates  the dominance of expansive effects due to negative $q$  over the contracting effects of inhomogeneous plasma. These observations suggest that the relative contraction effects on the shadow due to inhomogeneous  plasma is an interplay of  the plasma profile considered, the theory of gravity considered and the mass and  distance of the black hole (importantly the distance-to-mass ratio).   
    
\end{itemize}

 \newpage
\subsubsection{Constraints on tidal charge considering homogeneous plasma profile 3\label{Constraints from density plots  for Sgr A* considering plasma profile 3}}
\begin{figure}[h!]
    \centering
    \begin{subfigure}[b]{0.45\textwidth}
        \centering
        \includegraphics[width=\textwidth]{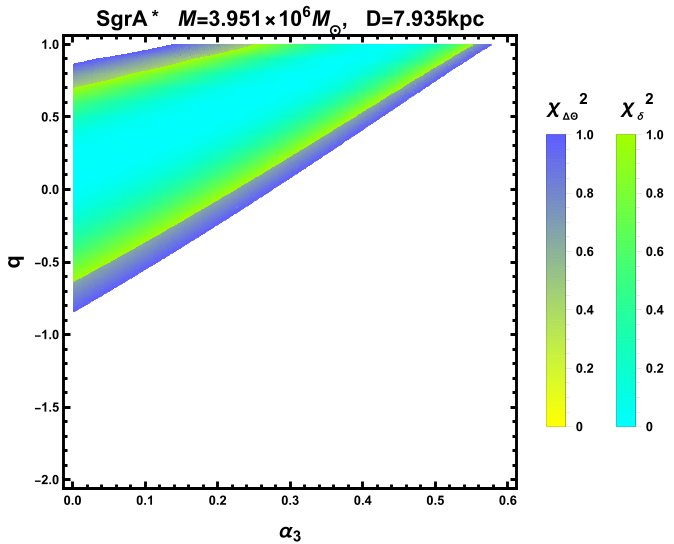} 
        \caption{Keck team \label{homogeneous_profile_combined_density_plot_SgrA_3.951.jpg}}
        
    \end{subfigure}
    \hfill
    \begin{subfigure}[b]{0.45\textwidth}
        \centering
        \includegraphics[width=\textwidth]{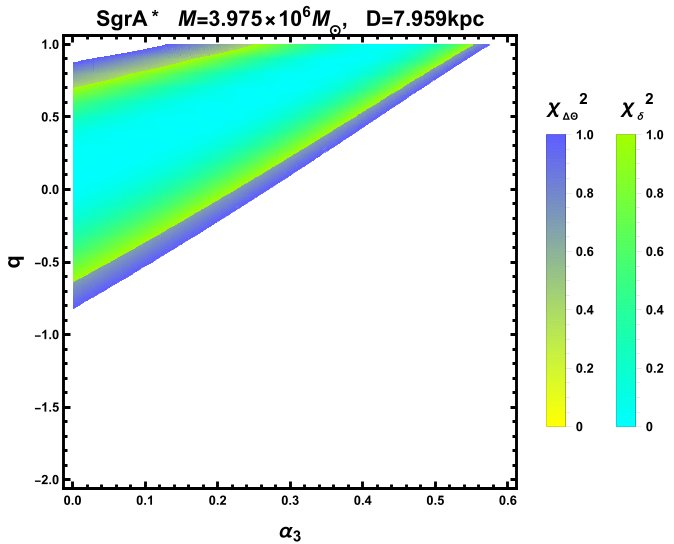} 
        \caption{Keck team \label{homogeneous_profile_combined_density_plot_SgrA_3.975.jpg}}
            \end{subfigure}
\vspace{0.5cm}
\begin{subfigure}[b]{0.45\textwidth}
        \centering
        \includegraphics[width=\textwidth]{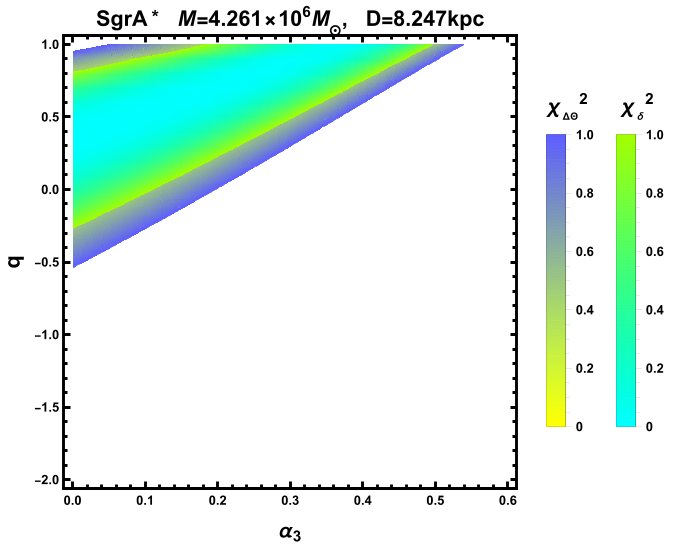} 
        \caption{GRAVITY collaboration \label{homogeneous_profile_combined_density_plot_SgrA_4.261.jpg}}
        
    \end{subfigure}
    \hfill
    \begin{subfigure}[b]{0.45\textwidth}
        \centering
        \includegraphics[width=\textwidth]{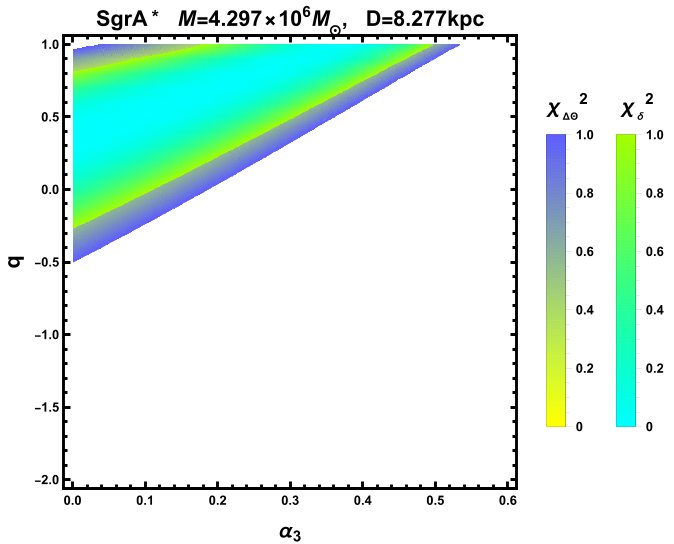} 
        \caption{GRAVITY Collaboration \label{homogeneous_profile_combined_density_plot_SgrA_4.297}}
        
    \end{subfigure}
    \caption{The  \cref{homogeneous_profile_combined_density_plot_SgrA_3.951.jpg,homogeneous_profile_combined_density_plot_SgrA_3.975.jpg,homogeneous_profile_combined_density_plot_SgrA_4.261.jpg,homogeneous_profile_combined_density_plot_SgrA_4.297} represent combined density plots of the parameter space constrained by the EHT estimate of shadow angular diameter and  Schwarzschild deviation parameter  for both the mass and distance estimates by the Keck  team and  the GRAVITY collaboration, respectively.  All the plots are computed considering plasma profile 1 and taking $\theta_i=46^\circ$. The overlapping region represents the allowed parameter space based on the error bars of both $\Delta\Theta_{obs}$  and $\delta_{sh}$.  \label{sgra_density_plots_profile3}}
\end{figure}

For the sake of completeness, we also discuss the case of homogeneous plasma for Sgr A*.  In \cref{sgra_density_plots_profile3}, the combined density plots of $\chi^2_{\Delta\Theta}$ and $\chi^2_{\delta}$ for Sgr A* are shown.   \cref{homogeneous_profile_combined_density_plot_SgrA_3.951.jpg,homogeneous_profile_combined_density_plot_SgrA_3.975.jpg} are the combined density plots for mass and distance estimates reported by the Keck team. \cref{homogeneous_profile_combined_density_plot_SgrA_4.261.jpg,homogeneous_profile_combined_density_plot_SgrA_4.297} are the combined density plots for mass and distance estimates reported by the GRAVITY collaboration.  We observe the following features from the density plots:
\begin{itemize}
    \item For all \cref{homogeneous_profile_combined_density_plot_SgrA_3.951.jpg,homogeneous_profile_combined_density_plot_SgrA_3.975.jpg,homogeneous_profile_combined_density_plot_SgrA_4.261.jpg,homogeneous_profile_combined_density_plot_SgrA_4.297}, the $\delta_{sh}$ gives more tighter constraints compared to $\Delta\Theta_{obs}$ as seen in \cref{sgra_density_plots_profile1,sgra_density_plots_profile2}. 
    \item When taking into account homogeneous plasma, the allowed range of $q$ varies with variation  of plasma parameter $\alpha_3$. From the density plots of  \cref{sgra_density_plots_profile3} , we observe that contrary to the case of inhomogeneous plasma profiles 1 and 2, the allowed upper and lower bounds of $q$ increase with $\alpha_3$.  This was also observed in the case of M87* when homogeneous plasma was considered (refer \cref{m87_density_plot profile3}).
    \item The EHT estimate for $\delta_{sh}$ and $\Delta\Theta_{obs}$ restrict the upper bound of plasma parameter $\alpha_3$. From \cref{homogeneous_profile_combined_density_plot_SgrA_3.951.jpg,homogeneous_profile_combined_density_plot_SgrA_3.975.jpg},  we observe that when mass and distance estimates of the Keck team are considered, $\alpha_3\gtrsim0.58$ is ruled out outside $1-\sigma$ of $\Delta\Theta_{obs}$. When $\delta_{sh}$ is considered the upper bound further reduces to $\alpha_3\gtrsim0.56$. In the case of mass and distance reported by the GRAVITY collaboration, $\alpha_3\gtrsim0.54$ is ruled out outside $1-\sigma $ of  $\Delta\Theta_{obs}$ and when $\delta_{sh }$ is considered, $\alpha_3\gtrsim0.5$ is ruled out outside $1-\sigma$ (refer \cref{homogeneous_profile_combined_density_plot_SgrA_4.261.jpg}).
    \item We further observe from \cref{sgra_density_plots_profile3} that, the difference between the allowed upper bounds of $\alpha_3$  for $\Delta\Theta_{obs }$ and $\delta_{sh}$ decreases as $q$ becomes more and more positive. This was also observed for M87* (refer \cref{m87_density_plot profile3}). This may be attributed to the interplay of the increasing effect of $\alpha_3$ and decreasing effect of positive tidal charge on the shadow.
\end{itemize}

As was observed in the case of M87* (refer  \cref{Constraints from density plots  for M87* considering plasma profile 1,Constraints from density plots  for M87* considering plasma profile 2,Constraints from density plots  for M87* considering plasma profile 3}),  the EHT estimates for angular diameter and Schwarzschild deviation parameter for Sgr A* are able to  put strong constraints on the tidal charge $q$ and plasma parameter $\alpha$ for plasma profiles 1, 2, and 3. However, contrary to the case of M87*, we observe that the EHT estimate for $\delta_{sh}$ imposed stronger constraints on the parameter space compared $\Delta\Theta_{obs}$,  for the case of Sgr A*. { Further, if we consider \cref{m87_density_plot profile1}, \cref{m87_density_plot profile2}, \cref{sgra_density_plots_profile1}, \cref{sgra_density_plots_profile2} we note that the observationally favored parameter space 
(the shaded region) spanned by $q$ and $\alpha$ shifts increasingly towards a negative tidal charge scenario when the plasma density (proportional to $\alpha_i$) is increased. The converse happens for the homogeneous plasma case (\cref{m87_density_plot profile3} and \cref{sgra_density_plots_profile3}). Thus, plasma can affect the shadow size and shape 
(\cref{Plots showing effect of plasma profile 1 on shadow of braneworld black hole}, \cref{Plots showing effect of plasma profile 2 on shadow of braneworld black hole} and \cref{Plots showing effect of homogeneous plasma on shadow of braneworld black hole}) provided its density is high. However, for low density plasma (with small $\alpha_i$) the shadow size and shape is predominantly determined by the background metric and in that case it really does not matter whether the low density plasma is inhomogeneous or homogeneous. Thus, if the photon encounters a low density plasma distribution ($\alpha_i<<1$) which is homogeneous or inhomogeneous, we will not be able to discern its signatures from the shadow size. Thus, the observationally favored values of $q$ for a small homogeneous plasma parameter e.g. $\alpha_3\sim 0.001$ and a small inhomogeneous plasma parameter are nearly the same (one may compare \cref{m87_density_plot profile1}, \cref{m87_density_plot profile2} with \cref{m87_density_plot profile3} for M87* and \cref{sgra_density_plots_profile1}, \cref{sgra_density_plots_profile2} with \cref{sgra_density_plots_profile3} for Sgr A*). Even for a higher plasma density, the constraints on $q$ do not vary much based on the plasma distribution, since the homogeneous plasma parameter $\alpha_3$ cannot exceed unity (to  enable light propagation through the plasma) and if we take e.g. $\alpha_3\sim 0.2$, both positive and negative $q$ (see,  for e.g., \cref {homogeneous_profile_combined_density_plot_M87_6.2.jpg}) are allowed. This allowed range of $q$ does not change drastically if an inhomogeneous plasma is considered, say, e.g., $\alpha_1, \alpha_2 \sim 0.2$ (\cref{r_profile_combined_density_plot_M87_6.2.jpg,theta_profile_combined_density_plot_M87_6.2.jpg} ). }

As was done in the case of M87*,   we can impose further strong constraints on the plasma parameter and the tidal charge by combining the EHT estimates of $\Delta\Theta_{obs}$ and $\delta_{sh}$ with the plasma density estimates for Sgr A* obtained from previous studies. The electron number density of Sgr A* varies based on the region considered; with the maximum number density  estimated to be of the order of $n_e\sim 10^5$cm$^{-3}$ in \citep{Ferriere}.  For   $r=5 r_g$ and $\theta=\pi/2$ we can estimate the plasma parameters $\alpha_1, \alpha_2$ and $\alpha_3$ using \cref{alpha1 and ne relation,alpha2 and ne relation,alpha3 and ne relation} for profile 1, 2 and 3, respectively.  Setting  $n_e=10^4$ cm$^{-3}, r=5 r_g$ and $\theta=\pi/2$ then using  \cref{alpha1 and ne relation,alpha2 and ne relation,alpha3 and ne relation}, we obtain $\alpha_1\sim 1.7 \times 10^{-11}$,  $\alpha_2\sim 1.2\times 10^{-11}$  and $\alpha_3\sim1.5\times10^{-12}$, respectively. 
Thus, we observe that the electron density estimate rules out very high values of $\alpha$. This implies that the background has a dominant effect on the shadow of Sgr A*.

Furthermore, we recall from \cref{perlick mdot alpha} for the case of profile 1,
$$  \alpha_1=\frac{e^2\ \dot{M}\ c^3}{\sqrt{2} m_e m_p \omega^2_0 G^2 M^2}$$
The above equation allows us to estimate the plasma parameter using previously reported estimates of $\dot{M}$ for Sgr A* { ($\omega_0$ is taken to be 230 GHz)}.  In the table below, we report estimates for $\alpha_1$ from $\dot{M}$ (reported from previous estimates).
\begin{table}[h!]
\centering
\begin{tabular}{|c|c|}
\hline
 {$\dot{M} $} &  {$\alpha_1$} \\
\hline
$ {\rm 10^{-7} M_\odot yr^{-1}-10^{-9} M_\odot yr^{-1}}$ \citep{Yoon:2020yew} & $2.95\times 10^{-8} -2.95\times 10^{-10}$  \\
\hline
$\sim {\rm 10^{-8} M_\odot yr^{-1}} $ \citep{Bower:2018wsw} &  $2.95\times 10^{-9}$ \\
\hline
$ {\rm 2\times 10^{-7} M_\odot yr^{-1}-2\times 10^{-9} M_\odot yr^{-1}}$ \citep{Marrone:2006vu}  & $5.89\times 10^{-8} -5.89\times 10^{-10}$ \\ \hline
$\lesssim {\rm 10^{-5} M_\odot yr^{-1}}$ \citep{Quataert:1999ng} & $\lesssim 2.95\times 10^{-6}$  \\
\hline
\end{tabular}
\caption{Estimates for $\alpha_1$ from the accretion rate of Sgr A*.}
\label{SgrA}
\end{table}
Considering the estimate of plasma parameters from the number density and \cref{SgrA}, we observe that  $\alpha\sim0$ is preferred  for the case of Sgr A*. This indicates, similar to the case of M87*, that for Sgr A* the background metric plays the dominant role in determining the size of the shadow. Thus, considering estimates of $\Delta\Theta_{obs},\delta_{sh}$ and $n_e$, the constraints obtained on $q$  for $\alpha\sim0$ are important for both M87* and Sgr A*.

\section{Conclusion\label{conclusion}}
In the present work, our goal is to investigate the  impact of the plasma environment on the shadow of a black hole in braneworld gravity.  In particular, we study the  impact of inhomogeneous and homogeneous plasma environments in modifying the shadow of braneworld black hole. When the plasma environment around the braneworld black hole is inhomogeneous, we observe that the size of the black hole shadow decreases as the density of the inhomogeneous plasma  increases. The contrary occurs when the surrounding plasma environment is homogeneous. Additionally, we observe that the shadow tends to become more circular as  the density of the inhomogeneous plasma increases, irrespective of the spin $a$ of the black hole. However,  the homogeneous plasma environment does not affect the shape of the black hole shadow, irrespective of its spin, although it increases its size.     
 {Such a study motivates us to search for signatures of braneworld gravity in the presence of plasma  using EHT estimates of the angular diameter of the shadow $\Delta\Theta_{obs}$ and the Schwarzschild deviation parameter $\delta_{sh,EHT}$ for M87* and Sgr A*.} In order to do this, we assumed M87* and Sgr A*  {to be}  rotating braneworld black holes with tidal charge $q$  and spin $a$ surrounded by plasma.  We then  obtained constraints on the tidal charge and plasma parameter by using the observables $\Delta\Theta_{obs}$ and $\delta_{sh}$ and estimates of $n_e$ for M87* and Sgr A*.  In order to arrive at the constraints, we used previously reported  mass, distance and angle of inclination  {estimates} for M87* and Sgr A* while  {considering} both inhomogeneous and homogeneous plasma environments.  {For the inhomogeneous plasma environment described by profile 1, the electron number density falls off as $n_e\sim r^{-3/2}$ \citep{shapiro1974accretion}, which has also been considered previously \citep{EventHorizonTelescope:2021srq,EventHorizonTelescope:2019pgp}.}

We observed that in the case of M87*, $\Delta\Theta_{obs}$  imposed more stringent constraints on the plasma parameter $\alpha$ and tidal charge $q$ compared to the Schwarzschild deviation parameter $\delta_{sh, EHT}$. Our analysis reveals that as the density of the inhomogeneous plasma environments (manifested through an increase in the plasma parameter) are increased, the allowed bounds on the tidal charge (based on $\Delta\Theta_{obs}$) decreases, thereby favouring a negative tidal charge. We observe the opposite for the homogeneous plasma scenario. Interestingly, the EHT estimate of  $\Delta\Theta_{obs}$ for M87* is able to impose tighter bounds on the plasma parameters compared to the theoretically allowed bounds
based on the light propagation condition for profiles 1,2 and 3 ( \cref{alpha max}).
Assuming that the tidal charge cannot be more negative than $q\sim -2$ (based on previous studies \citep{Bhattacharya:2016naa}), we report that $\alpha_1\gtrsim10$, $\alpha_2\gtrsim17$  and $\alpha_3\gtrsim0.56$ are ruled out outside $1-\sigma$  for profiles 1, 2 and 3, respectively, when the mass estimate from the stellar dynamics studies is considered to reproduce $\Delta\Theta_{obs}$.  
Using estimates of electron number density $n_e$ near the event horizon of M87* from the EHT analyses ($n_e\sim10^4$–$10^7~\mathrm{cm^{-3}}$ at $r\sim5r_g$) and the reported accretion rate $\dot{M}\sim10^{-4}$–$10^{-3}~M_\odot~\mathrm{yr^{-1}}$, the corresponding plasma parameter obtained from \cref{alpha1 and ne relation,alpha2 and ne relation,alpha3 and ne relation} lies in the range $\alpha\sim10^{-10}$–$10^{-7}$. These values are orders of magnitude smaller than the upper bounds derived from the shadow fits ($\alpha_1\gtrsim10$, $\alpha_2\gtrsim17$, $\alpha_3\gtrsim0.56$), implying that the plasma contribution is negligible on the observed size of the shadow, i.e., 
the spacetime geometry dominates over plasma in shaping the size of the observed shadow. This corroborates with the findings of the EHT collaboration \citep{EventHorizonTelescope:2025vum,EventHorizonTelescope:2024dhe} which report the persistent shadow of M87* when monitored during 2017, 2018 and 2021. 
Since the electron number density estimate leads to $\alpha\sim 0$, the allowed range of tidal charge is found to be $-1.15\lesssim q\lesssim0.45$ for inhomogeneous plasma environments (profiles 1 and 2), considering  mass estimate from stellar dynamics studies. When the mass estimate from gas dynamics studies is considered along with $\alpha\sim 0$,  we do not reproduce  $\Delta\Theta_{obs}$ and $\delta_{sh, EHT}$ within the $1-\sigma$ estimates of the EHT, for inhomogeneous plasma environment surrounding M87*. It is important to note that, even with the mass estimate from gas dynamics studies we may be able to explain the observed shadow size in the absence of plasma for some $q<-2$. This is because of the expanding effect of negative tidal charge on the shadow size which compensates for the relatively small mass estimated from gas dynamics studies.   However, very high magnitudes of negative tidal charges have been found to be undesirable in previous studies \citep{Bhattacharya:2016naa}.

In the case of Sgr A*,  we observed that  $\delta_{sh, EHT}$ imposed  stronger constraints on $q$ and $\alpha$ compared to $\Delta\Theta_{obs}$. As the inhomogeneous plasma density is increased (by increasing $\alpha$) the allowed bounds on $q$ within $1-\sigma$ of the EHT estimates shift towards a more negative tidal charge scenario. The opposite was observed when  {the homogeneous} plasma environment was considered. 
When mass and distance estimates from the Keck team and the GRAVITY collaboration are respectively considered, both the inhomogeneous plasma environments rule out $\alpha_1, \alpha_2 \gtrsim 11$ and $\alpha_1, \alpha_2 \gtrsim12$ (for $q\sim -2)$.  {When the} homogeneous plasma environment is considered, $\alpha_3\gtrsim0.54$ was ruled out outside $1-\sigma $ of $\delta_{sh, EHT}$ (for $q\sim -2)$.
When electron density estimates for Sgr~A* ($n_e\lesssim10^5~\mathrm{cm^{-3}}$) or accretion rate constraints ($\dot{M}\sim10^{-9}$–$10^{-7}~M_\odot~\mathrm{yr^{-1}}$) from previous studies are considered, the corresponding $\alpha$ values ($\alpha\sim10^{-11}$–$10^{-8}$) turn out to be  {negligible,} which again shows the dominance of the background metric on the size of the shadow for Sgr A* \citep{EventHorizonTelescope:2022wkp}.  For $\alpha\sim0$, the allowed range of tidal charge within $1-\sigma$ of $\delta_{sh, EHT}$ is found to be $-0.65\lesssim q\lesssim0.7$ (based on mass and distance estimated by the Keck team) and $-0.3\lesssim q\lesssim0.8$ (based on mass and distance estimated by the GRAVITY collaboration).

For both M87* and Sgr A*,  $\Delta\Theta_{obs}$ and $\delta_{sh,EHT}$  {fail} to impose constraints on the spin of the  {sources}, which can be attributed to the weak dependence of the shadow size on the BH spin at low inclination angles. Taken together, these results clarify that the EHT observables $\Delta\Theta_{\mathrm{obs}}$ and $\delta_{sh,EHT}$ do not, by themselves, provide evidence for metric dominance, as both are influenced by an interplay of geometry and plasma effects. Since M87* and Sgr A* are surrounded by a low density plasma, the background geometry plays a dominant role in determining the observed shadow size. However, for dense plasma environments, our study showed that the plasma can influence the shadow size and hence the constraints on tidal  {charge;} this may be important in the case of black holes with dense plasma environments\citep{Jiang:2019ztr}. 
 {Our}  work therefore highlights the importance of combining shadow observables with independent constraints on the accretion environment to disentangle geometric and plasma effects. Future EHT observations, multi-band imaging, and improved polarimetric modelling of accretion flows will be essential to provide more stringent constraints on the tidal charge, enabling robust tests of higher-dimensional gravity in realistic astrophysical environments.


\bibliography{references}
\bibliographystyle{utphys1}
\end{document}